\documentclass[letterpaper,twocolumn,10pt]{article}
\usepackage{usenix2019_v3}

\usepackage{tikz}
\usepackage{amsmath}
\usepackage{amsthm}
\usepackage{amssymb}
\usepackage{wasysym}
\usepackage[]{graphicx}
\graphicspath{{./Figure/}}
\usepackage{algorithm}
\usepackage[noend]{algpseudocode}
\usepackage{booktabs}
\usepackage{multirow} 
\usepackage{multicol}
\usepackage{makecell}
\usepackage{threeparttable}
\usepackage{xcolor}
\usepackage{enumitem}
\usepackage{fancyhdr}
\usepackage{hyperref}
\usepackage{listings}
\usepackage[algo2e]{algorithm2e} 
\usepackage{float}

\definecolor{codegreen}{rgb}{0,0.6,0}
\definecolor{codegray}{rgb}{0.5,0.5,0.5}
\lstdefinelanguage{Rust}{
    keywords=[1]{as, break, const, continue, crate, else, enum, extern, false, fn, for, if, impl, in, let, loop, match, mod, move, mut, pub, ref, return, self, static, struct, super, trait, true, type, unsafe, use, where, while, async, await, dyn},
    keywords=[2]{Self, Copy, Send, Sized, Sync, Drop, Fn, FnMut, FnOnce, Box, Vec, String, Option, Result, Some, None, Ok, Err, From, Into, Default},
    keywordstyle=[1]\color{codegreen}\bfseries,  
    keywordstyle=[2]\color{blue}\bfseries,  
    commentstyle=\color{codegray},  
    stringstyle=\color{red},  
    morecomment=[l][\color{gray}]{//}, 
    morecomment=[s][\color{gray}]{/*}{*/}, 
    morestring=[b]{"},
}
\usepackage[available]{usenixbadges}

\newcommand*\BC[1]{%
\begin{tikzpicture}[baseline=(C.base)]
\node[draw,circle,fill=black,inner sep=0.2pt](C) {\textcolor{white}{#1}};
\end{tikzpicture}}

\newcommand{\tyanalyzer}{{property graph constructor}\xspace}
\newcommand{\Tyanalyzer}{\textsf{Property Graph Constructor}\xspace}
\newcommand{\Bugdetector}{\textsf{Bug Detector}\xspace}
\newcommand{\bugdetector}{{bug detector}\xspace}
\newcommand{\bug}{type confusion bug\xspace}
\newcommand{\bugs}{type confusion bugs\xspace}

\newcommand{\bone}{misalignment\xspace}
\newcommand{\btwo}{inconsistent layout\xspace}
\newcommand{\bthree}{mismatched scope\xspace}
\newcommand{\Bone}{Misalignment\xspace}

\newcommand{\bdone}{misalignment detector\xspace}
\newcommand{\bdtwo}{inconsistent layout detector\xspace}
\newcommand{\bdthree}{mismatched scope detector\xspace}
\newcommand{\Bdone}{Misalignment Detector\xspace}
\newcommand{\Bdtwo}{Inconsistent Layout Detector\xspace}
\newcommand{\Bdthree}{Mismatched Scope Detector\xspace}

\newcommand{\TN}{\textsc{TypePulse}\xspace}

\newcommand{\pcg}{property graph\xspace}

\newcommand{\Analysisone}{Type Conversion Analysis\xspace}
\newcommand{\analysisone}{type conversion analysis\xspace}
\newcommand{\Analysistwo}{Pointer alias analysis\xspace}
\newcommand{\analysistwo}{pointer alias analysis\xspace}
\newcommand{\Checkone}{Type conversion check\xspace}
\newcommand{\checkone}{type conversion check\xspace}
\newcommand{\Checktwo}{Access check\xspace}
\newcommand{\checktwo}{access check\xspace}
\newcommand{\rs}{RUSTSEC\xspace}
\newcommand{\tc}{developer-enforced check\xspace}
\newcommand{\TC}{Developer-Enforced Check\xspace}

\def\Snospace~{\S{}}

\newenvironment{packeditemize}{
\begin{list}{$\bullet$}{
\setlength{\labelwidth}{6pt}
\setlength{\itemsep}{2pt}
\setlength{\leftmargin}{\labelwidth}
\addtolength{\leftmargin}{\labelsep}
\setlength{\parindent}{1pt}
\setlength{\listparindent}{\parindent}
\setlength{\parsep}{1pt}
\setlength{\topsep}{1pt}}}{\end{list}}

\begin{document}

\date{}

\title{\TN:~Detecting Type Confusion Bugs in Rust Programs}

\author{
{\rm Hung-Mao Chen$^*$, Xu He$^*$, Shu Wang$^*$$^\dag$, Xiaokuan Zhang$^*$, Kun Sun$^*$}\\
$^*$George Mason University\\
$^\dag$Palo Alto Networks
}
\maketitle

\pagenumbering{gobble}

\begin{abstract}
Rust supports type conversions and safe Rust guarantees the security of these conversions through robust static type checking and strict ownership guidelines. However, there are instances where programmers need to use unsafe Rust for certain type conversions, especially those involving pointers. Consequently, these conversions may cause severe memory corruption problems. Despite extensive research on type confusion bugs in C/C++, studies on type confusion bugs in Rust are still lacking. Also, due to Rust’s new features in the type system, existing solutions in C/C++ cannot be directly applied to Rust. In this paper, we develop a static analysis tool called \TN to detect three main categories of \bugs in Rust including misalignment, inconsistent layout, and mismatched scope. \TN first performs a {type conversion analysis} to {collect and determine trait bounds for type pairs}. Moreover, it performs a pointer alias analysis to resolve the alias relationship of pointers. {Following the integration of information into the \pcg}, it constructs type patterns and detects each type of bug in various conversion scenarios. We run \TN on the top 3,000 Rust packages and uncover {71} new \bugs, exceeding the total number of \bugs reported in RUSTSEC over the past five years. We have received {32} confirmations from developers, along with one CVE ID and six RUSTSEC IDs.
\end{abstract}
\section{Introduction}

Rust~\cite{RustLang} is an emerging programming language known by its strict enforcement of type safety and memory safety through compile-time checking, without sacrificing runtime performance. Since memory safety issues in unsafe languages such as C and C++ have been known to lead to catastrophic consequences, Rust has become an appealing solution to replace C and C++, and it has been adopted in major open-sourced projects such as the Linux kernel~\cite{LinuxKernel} and the Firefox browser~\cite{MozillaFirefox}. Recently, the White House also calls for adoption of memory-safe programming languages such as Rust to secure the cyberspace~\cite{whitehouse}.

Rust is fundamentally divided into two separate sub-languages: safe Rust and unsafe Rust~\cite{safeunsafe}. Safe Rust enforces strict compile-time checks to maintain memory safety and type safety. However, these checks can be excessively restrictive, blocking some essential but risky operations like accessing raw pointers. To address this, Rust introduces the {\tt unsafe} keyword for such situations. When employing unsafe Rust, it falls upon the programmer to uphold memory safety since the typical compile-time checks are circumvented.

Similar to traditional programming languages such as C/C++, Rust also supports type conversions~\cite{TypeConv22online}, where a variable is initially converted from type $A$ to type $B$  and subsequently accessed as type $B$. Safe Rust guarantees the security of these conversions through robust static type checking and strict ownership guidelines. Rust automatically infers the types of variables and expressions from their context and use. Also, the ownership feature helps ensure memory and concurrency safety by tracking the lifetime and borrowing at compile time~\cite{Ownership}. Nevertheless, there are instances where programmers need to use unsafe Rust for certain type conversions, especially those that involve pointers. For instance, since direct reference conversions are not allowed, conversions must be first made at the raw pointer level and then converted back to references using unsafe Rust. Consequently, these conversions may cause severe memory corruption problems similar to those found in C/C++~\cite{cve-2023-3079, cve-2023-4762, cve-2024-1939}. In the last five years, RUSTSEC~\cite{rustsec} has reported 32 type confusion bugs that can lead to various memory-safety issues such as data leaks, uninitialized memory, and Out-Of-Bounds (OOB) memory access.

Despite extensive research on type confusion bugs in C and C++~\cite{jeon2017hextype, haller2016typesan, duck2018effectivesan, lee2015type}, studies on type confusion bugs in Rust are still lacking due to three significant challenges. First, the conversion of data types between functions complicates type analysis, and this complexity cannot be addressed using traditional interprocedural analysis. For instance, when a type constructor function creates an instance of a type that is then handed off to another function for an unsafe type conversion, traditional interprocedural analysis fails to track the constructor function as it is absent from the conventional call graph~\cite{rupta}. Therefore, a new call graph is needed to identify this type of interprocedural type conversion.

{Second, predicting all possible concrete types that can replace a generic type in Rust is inherently challenging. Unlike C and C++, Rust uses trait bounds~\cite{Traitand70:online} to constrain generic types, ensuring they conform to specific behaviors and capabilities. It adds complexity since trait bounds can have implicit and recursive dependencies on other traits. Moreover, concrete types may encompass composite types like \texttt{struct}. As a result, a method of generic type resolution that adheres to the trait bounds is needed.}

Third, identifying \bugs requires establishing whether the pointer alias remains valid after type conversion, which is essential for our bug verification process. However, the ownership and lifetime features in Rust increases the complexity when undertaking pointer analysis. In C++, existing techniques primarily address the issue of pointer access via alias analysis~\cite{10.1145/3503222.3507770, 10301168, fan2020accelerating}. Nevertheless, traditional alias analysis needs modification to accommodate the pointer variable ownership and lifetime in Rust. For example, when ownership is transferred to another pointer or the pointer is automatically deallocated, the original pointer variable becomes invalid,
which cannot be detected by existing techniques.

To tackle these challenges, we develop \TN, a static analysis tool to detect \bugs in Rust applications. It consists of two main components, namely, \Tyanalyzer and \Bugdetector. By examining each function within the crate, \tyanalyzer creates a new call graph that helps identify the type constructor functions, addressing the first challenge. Property graph constructor performs \analysisone and \analysistwo. Type conversion analysis is conducted to collect type pairs (i.e., $\text{<}$source type, destination type$\text{>}$) and trait bounds for generic types through dependency resolution. It is employed to address the second challenge. Pointer alias analysis will construct a new alias graph, representing the alias relationship of the pointers. When ownership is transferred or dropped, \tyanalyzer updates the alias graph to reflect the node connections, which can solve the third challenge. As the final output, each function in the \pcg is associated with type pairs, trait bounds, and the pointer alias graph. The \pcg will be utilized by \bugdetector to analyze and verify the type confusion bugs.

Next, \bugdetector utilizes {the \pcg} to perform type conversion checks and access checks. First, type conversion checks are used to identify if the type conversion creates an invalid type pointer. Second, access checks will be performed to analyze if the invalid type pointer can be accessed. Via analyzing the type pairs and trait bounds, we summarize three patterns of type confusion bugs, namely, \bone, \btwo, and \bthree, to help locate invalid type pointers. When performing access checks, \bugdetector will first traverse the alias graph to determine if the invalid type pointer is accessible, then verify the absence of {\tc}, which is manually implemented by developers to prevent the type confusion bugs. After verification, the bug report will be delivered with the unsafe type conversion and type access highlighted.

We implement a prototype of \TN with 5249 lines of Rust. To assess \TN's effectiveness in bug identification, we first execute it on the \bugs reported to RUSTSEC from 2019 to 2024. The findings show that \TN can successfully identify all reported \bugs. Next, we perform a large-scale study by running \TN on the top 3,000 popular packages ranked on \url{crates.io} and \url{GitHub}. We detect 71 new \bugs and report all of them to their package developers. We receive {32} confirmations on the reported bugs at the time of writing. The new bugs occur in many high-profile repositories. For example, we demonstrate that the \bug within the \texttt{pprof} package can cause the downstream applications to crash (e.g., GreptimeDB with 4.2k stars on GitHub).  \looseness=-1

\vspace{0.1in}
\noindent{\bf Contributions.}
This paper makes the following contributions: 
\begin{packeditemize}
    \item We analyze all \bugs in RUSTSEC in the last five years and identify the three most prevalent categories of \bugs, namely, misalignment, inconsistent layout, and mismatched scope.
    \item We design and implement the first static analysis tool (\TN) to detect \bugs in Rust, {addressing the challenges of interprocedural type conversion, generic type resolution, and alias analysis} due to the unique features of Rust.
    \item We evaluate \TN on the top 3,000 Rust packages and identify {71} new \bugs which we have manually confirmed, surpassing the total number of bugs reported in RUSTSEC in the last five years. We also run \TN on existing \bugs and it achieves 100\% accuracy, demonstrating the robustness of \TN. 
\end{packeditemize}

\section{Background}
\label{sec:bg}

\subsection{Rust Basics}

\noindent{\bf Generic Types and Traits.}
Rust provides the flexibility of code reuse and type safety with generic types and traits \cite{GenericD71:online, traits}. By writing code in a type-agnostic manner, generic types allow functions, methods, or data structures (e.g., struct) to operate on multiple types, which are represented by the placeholder (e.g., \texttt{T}, \texttt{U}). Since the generic types will be replaced by the actual types at the compile-time (i.e., monomorphization), it can prevent the \bugs at run-time. In addition, traits can be used to specify the constraints on generic types. By applying a trait to a generic type, Rust ensures that the type used to replace the generic type should also implement the required methods or characteristics defined by the trait. For example, the function \texttt{fn display<T: Copy>(input: T)} ensures that only the type implementing the \texttt{Copy} trait can be initialized as the argument \texttt{input}.

\vspace{0.05in}
\noindent{\bf Safe vs. Unsafe Rust.}
The central concept of safe Rust is to confirm memory {\it ownership} during compilation, where the compiler checks both the access and the {\it lifetime} of memory-allocated objects (or values). Moreover, safe Rust permits the {\it borrowing} of a value (i.e., making a reference to it) throughout the lifetime of the owner variable. In contrast, \texttt{unsafe} is used to highlight code segments that perform tasks that are not ensured by the compiler, placing the responsibility on developers to prevent memory safety issues. In Rust, there are five specific actions necessitating the \texttt{unsafe} keyword~\cite{unsafeRu19online}: dereferencing raw pointers, invoking unsafe functions, altering or accessing mutable static variables, defining unsafe traits, and executing inline assembly. Each of these actions could breach Rust's safety assurances. Unsafe Rust is crucial because it allows developers to interface with low-level system APIs, libraries written in other languages, or hardware directly.

\vspace{0.05in}
\noindent{\bf Undefined Behaviors.}
Undefined behavior (UB) refers to the program whose outcome is not prescribed by the language's specification, which means that the language standard does not define what should happen if the UB occurs. In most cases, the result can only be decided by hardware and architectures, leading to inconsistent consequences in different environments. The outcomes of UB are unpredictable, ranging from security vulnerabilities to incorrect compiler optimization and code generation. The backend of the compiler might perform the optimization based on the assumption that the UB will not occur. Therefore, we should prevent UB from happening. Rust clearly defines scenarios that might trigger undefined behaviors~\cite{ubreference}, such as dereferencing null pointers, accessing out-of-bounds array elements, and data races when mutating shared data without synchronization. The design of \texttt{unsafe} help us narrow down the culprit of UB to the code related to unsafe code.

\subsection{Type Conversion in Rust}
Type conversion from the source type (\texttt{src\_ty}) to the destination type (\texttt{dst\_ty}) consists of two steps: 
\BC{1} {\it Conversion}, which involves altering or reinterpreting the bit pattern of a variable from one type (\texttt{src\_ty}) to a new type (\texttt{dst\_ty}), and \BC{2} {\it Access}, which involves accessing the variable as the new type (\texttt{dst\_ty}).
\texttt{rustc} limits developers to performing only explicit type conversions to maintain safety with compile-time verifications. These conversions can be implemented through type casting, transmute operations, and traits~\cite{traits}. Given that traits are typically handled by casting and transmute methods, our discussion will focus solely on casting and transmute operations.

\lstdefinestyle{lst}{
    float=tp,
    floatplacement=tbp,
    numbers=left, 
    numberstyle=\scriptsize, 
    numbersep = 5pt,
    framexleftmargin = 0in,
    framexrightmargin = 0in,
    xleftmargin = 0.18in,
    xrightmargin = 0.1in,
    basicstyle=\ttfamily\scriptsize, 
    frame=lines,
    showtabs=true,
    showspaces=true,
    showstringspaces=false,
    literate={\ }{{\ }}1,
}

\begin{lstlisting}[
language=rust, 
style=lst,
caption=Type conversion between pointers in \texttt{unsafe} code.,
label=listing:motivation1,
mathescape=true]
fn main() {
  let source_ty: u8 = 1;
  // compile error: non-primitive cast
  let dest_ty = &source_ty as &u32;  
  // Alternative 1: as
  let tmp_ty= &source_ty as *const u8 as *const u32;
  let dest_ty = unsafe {&*tmp_ty};
  // Alternative 2: transmute
  let dest_ty = unsafe { 
    transmute::<&u8, &u32>(&source_ty) 
  };
}
\end{lstlisting}

\vspace{0.05in}
\noindent{\bf Casting Operation.}
The type-casting operation depends on the keyword \texttt{as}, which is mainly used for secure and direct type conversions, including conversions between basic data types and raw pointers. The use of \texttt{as} generally involves straightforward bit manipulations or adjustments in values. For example, when an \texttt{f32} is converted to an \texttt{i32}, the fractional component of the floating point number is removed. Consequently, using \texttt{as} can result in data truncation or loss. It's important to note that under the stringent regulations established by \texttt{rustc}, \texttt{as} can be employed in the {\tt safe} code.

\vspace{0.05in}
\noindent{\bf Transmute Operation.}
Compared to \texttt{as}, the \texttt{transmute} function facilitates more intricate and dangerous transformations. Essentially, \texttt{transmute} performs a bitwise copy from one type to another without altering the bit pattern of the value. However, it modifies the interpretation of these bits by \texttt{rustc}. For instance, it allows for the direct conversion of an \texttt{i32}'s bit pattern to an \texttt{f32}, despite their fundamentally different representations. The validity of the original bit pattern in the new type is not assured, making \texttt{transmute} extremely risky and prone to causing undefined behaviors. Consequently, \texttt{transmute} necessitates the use of \texttt{unsafe} code.

\vspace{0.05in}
\noindent{\bf Type Conversion between Pointers.}
In Rust, pointer-type conversions can be achieved through casting and the use of transmute operations. However, Rust imposes various restrictions depending on the pointer types involved, meaning that some conversions are safely handled by Safe Rust, while others require the use of the \texttt{unsafe} keyword. For example, in Listing~\ref{listing:motivation1}, Safe Rust prohibits direct conversion between reference types (line 3), forcing developers to resort to two methods within \texttt{unsafe} code. The first method involves converting the reference of the original type to a raw pointer, followed by its conversion to another raw pointer (line 5). To acquire a reference of the new type, developers must first dereference the raw pointer and then form a new reference (line 6). As dereferencing a raw pointer is not allowed in Safe Rust, the use of \texttt{unsafe} code becomes unavoidable. Alternatively, the \texttt{transmute} function can be used directly to convert references (line 9), which must be used with the \texttt{unsafe} keyword due to its inherent risks. Such conversions in Unsafe Rust are prone to type conversion errors because the memory address remains the same for both pointers~(\texttt{\&src\_ty} and \texttt{\&dst\_ty}), while only the interpretation of the type changes.

\begin{table}[t] \label{rustsec-table}
\centering
\caption{The details of 32 reported Type Confusion Bugs on RustSec advisories (2019-2024).} \label{rustsec-details}
\vspace{0.05in}
\footnotesize
\begin{tabular}{c c c c c}
\toprule
  Year   & Type I & Type II  & Type III & Others \\ \midrule
2019 & 2019-0035 & - & 2019-0028 & - \\ \hline
2020 & 
\begin{tabular}[c]{@{}c@{}}2020-0035\\ 2020-0050\end{tabular} & 
\begin{tabular}[c]{@{}c@{}}2020-0029\\ 2020-0078\\ 2020-0079\\ 2020-0080\\ 2020-0081\\ \end{tabular} & 
\begin{tabular}[c]{@{}c@{}}2020-0029\\ 2020-0165 \end{tabular} &
\begin{tabular}[c]{@{}c@{}}2020-36317\\ 2020-0073\\ 2020-0164\end{tabular} \\ \hline
2021 & 
\begin{tabular}[c]{@{}c@{}}2021-0120\\ 2021-0121\\ 2021-0145\end{tabular} & 
\begin{tabular}[c]{@{}c@{}}2021-0021\\ 2021-0035\end{tabular} & 
\begin{tabular}[c]{@{}c@{}}2021-0019\\ 2021-0089\end{tabular} & 2021-0044 \\ \hline
2022 & 2022-0041 & 
\begin{tabular}[c]{@{}c@{}}2022-0052\\ 2022-0074\end{tabular} & 
2022-0092 & 
\begin{tabular}[c]{@{}c@{}}2022-0034\\ 2022-0078\end{tabular} \\ \hline
2023 & - & - & 
\begin{tabular}[c]{@{}c@{}}2023-0015\\ 2023-0055\end{tabular}  & - \\ \hline
2024 & - & 2024-0347 & 
\begin{tabular}[c]{@{}c@{}}2024-0001\end{tabular}  & - \\ 
\bottomrule
\end{tabular}
\end{table}

\section{Overview}

\subsection{Motivating Examples}
\label{sec:ty-bug}
We investigate all bug reports related to \bug{s} in the RUSTSEC advisories in the last five years~\cite{rustsec}. There are 32 reports but only 26 \bugs; The remaining six bugs are out of consideration because they are related to other memory safety problems, such as Use-After-Free, which can be handled by existing tools~\cite{Qin2020ReplicationPF, Yechan2021Rudra, Zhuohua2021MirChecker, safedrop}. The Bug IDs are listed in \autoref{rustsec-details}. We categorize the 26 bugs into three types based on their root causes, which are \bone, \btwo, and \bthree. All 26 bugs are related to pointer type conversion (i.e., references and raw pointers) in Unsafe blocks. In this section, we introduce the three bug types and their security impacts.

\vspace{0.05in}
\noindent\textbf{Type I: \Bone Bug.} The first type of bug occurs when the type is converted to another type leading to alignment violation. The alignment of a type specifies the valid memory address at which the type should be stored.  Given the alignment value as \texttt{n}, the type must be stored only at the address of a multiple of \texttt{n}. Some types have a fixed alignment regardless of the target architectures, while others could be platform-specific.  For example, the type of \texttt{i32} has both 4-byte alignments on the 32-bit or 64-bit target. In contrast, the types of \texttt{usize} and \texttt{isize} are aligned to 4 bytes on the 32-bit target and 8 bytes on the 64-bit target. The alignment requirement can be easily violated with pointer-type conversion since two pointers remain in the same memory address. Although the memory address is aligned for \texttt{src\_ty}, it might not be aligned for \texttt{dst\_ty} if not handled carefully. In the Listing \ref{listing:motivation-type1}, the method \texttt{fill\_bytes} allows the slice of \texttt{u8} to be cast onto the slice of \texttt{u32} (line 9). Since \texttt{u8} is aligned to 1 byte, the slice of \texttt{dest} can be stored at the arbitrary memory address. When it turns out to be accessed as \texttt{u32}, it is not guaranteed that the memory address can be multiple of 4 since \texttt{u32} is aligned to 4 bytes, leading to the misaligned pointer dereference. Note that developers of \texttt{rand\_core} consider that the issue could be avoided by limiting the target architectures to \texttt{x86} or \texttt{x86\_64} only since these architectures are designed to tolerate misaligned memory access. However, the alignment requirement is enforced by the compiler instead of these target architectures. Once the Rust compiler verifies that safe code adheres to alignment rules, it generates optimized machine code based on this assurance. However, if unsafe code violates these rules, it can cause undefined behavior or crash the program.
\looseness=-1

\lstdefinestyle{lst}{
    float=tp,
    floatplacement=tbp,
    numbers=left, 
    numberstyle=\scriptsize, 
    numbersep = 5pt,
    framexleftmargin = 0in,
    framexrightmargin = 0in,
    xleftmargin = 0.18in,
    xrightmargin = 0.1in,
    basicstyle=\ttfamily\scriptsize, 
    frame=lines,
    showtabs=true,
    showspaces=true,
    showstringspaces=false,
    literate={\ }{{\ }}1,
}

\begin{lstlisting}[
language=rust, 
style=lst,
caption=A misalignment bug in \texttt{rand\_core} that casts bytes slices to integer slices (RUSTSEC-2019-0035 \cite{rs-2019-0035}).,
label=listing:motivation-type1,
mathescape=true]
#[cfg(any(target_arch = "x86",
  target_arch = "x86_64"))]
fn fill_bytes(&mut self, dest: &mut [u8]) {
  // ...

  while filled < end_direct {
    let dest_u32: &mut R::Results = unsafe { 
      &mut *(dest[filled..].as_mut_ptr() as 
      *mut <R as BlockRngCore>::Results) };
    self.core.generate(dest_u32);
    filled += self.results.as_ref().len() * 4;
    self.index = self.results.as_ref().len();
  }
  // ...
}
\end{lstlisting}

\vspace{0.05in}
\noindent\textbf{Type II: Inconsistent Layout Bug.} \label{motiv:bugII} The second type of bug occurs when \texttt{src\_ty} and \texttt{dst\_ty} have different memory layouts. 
In Listing \ref{listing:motivation-type2}, the method \texttt{as\_ref} allows casting between \texttt{Table} and \texttt{TableSlice} and returns the reference to the new type (line 20). 
However, when the \texttt{struct} type in Rust inherits the default representation (e.g. \texttt{repr(Rust)}), the compiler may reorder the memory layout, such as the fields of \texttt{struct}. 
The results of GDB show that after the raw pointer to \texttt{Table} is converted to the \texttt{TableSlice}, the fields \texttt{rows} of \texttt{Table} and one of \texttt{TableSlice} point to different memory addresses (line 26 and line 28), leading to inconsistent lengths of \texttt{rows} (line 27 and line 29).
It could impact applications that rely on the value (e.g., \texttt{rows.length}). 
For example, when applications plan to print all the data stored in table format to the terminal, the API (\texttt{TableSlice::print\_tty}) converts \texttt{Table} to \texttt{TableSlice} first and iterates the data stored in \texttt{rows}. 
Since iterating slice relies on the length (line 29) while the number of elements is actually only one (line 27), printing table leads to invalid memory access and segmentation fault, which has been reported in RUSTSEC-2022-0074.

\begin{lstlisting}[
language=rust, 
style=lst,
caption=An inconsistent layout bug in \texttt{prettytable-rs} that casts a \texttt{\&Vec} to \texttt{\&[T]} (RUSTSEC-2022-0074 \cite{rs-2022-0074}).,
label=listing:motivation-type2,
mathescape=false]
pub struct Table {
  format: Box<TableFormat>,
  titles: Box<Option<Row>>,
  rows: Vec<Row>,
}

pub struct TableSlice<'a> {
  format: &'a TableFormat,
  titles: &'a Option<Row>,
  rows: &'a [Row],
}

impl<'a> AsRef<TableSlice<'a>> for Table
fn as_ref(&self) -> &TableSlice<'a> {
  unsafe {
    let s = &mut *(
      (self as *const Table) as *mut Table
    );
    s.rows.shrink_to_fit();
    &*(self as *const Table as *const TableSlice<'a>)
  }
}

// From GDB results
// $8 as &Table, $7 as &TableSlice
p &$8.rows       // 0x7ff..82f0
p &$8.rows.len   // 1
p &$7.rows       // 0x7ff..82e0
p $7.rows.length // 93825009397280!
\end{lstlisting}

\vspace{0.05in}
\noindent\textbf{Type III: Mismatched Scope Bug.} The third type of bug occurs when we break the invariant by creating an invalid bit pattern or modifying the mutability of types. In Listing \ref{listing:motivation-type-3}, the trait \texttt{ComponentBytes} is designed to provide a method \texttt{as\_bytes\_mut} to modify the type \texttt{T} as byte slice. The type \texttt{T} could be any type that implements the traits \texttt{Copy}, \texttt{Send}, \texttt{Sync}, and lifetime bound \texttt{static}. However, the trait and lifetime bounds here cannot prevent the issues caused by the problematic type conversion implemented in \texttt{as\_bytes\_mut}. It allows casting between mutable raw pointers (\texttt{slice.as\_mut\_ptr() as *mut u8}) to create an invalid state for types since two pointers are pointing to overlapping memory. Safe Rust enforces aliasing rules, where mutable and immutable references can not exist simultaneously, while unsafe Rust allows the rule to be bypassed, as shown in the exploit (line 21 - 27). The attacker creates an immutable reference pointing to the static string as the type \texttt{T}. With \texttt{as\_bytes\_mut}, the mutable raw pointer to the slice of string casts to the mutable raw pointer of \texttt{u8} type. Since the function returns a mutable reference to a slice of \texttt{u8}, the attacker is allowed to modify any values in the slice of \texttt{u8}. However, the mutable reference \texttt{bytes} and immutable reference \texttt{component} point to the same data, breaking the aliasing rules of safe Rust. While the attacker modifies the value in \texttt{bytes}, he also changes the value of \texttt{component}, which should not be mutated originally. One security consequence of modifying immutable data is data races. In a multi-threaded environment, if the immutable object can be modified through a mutable reference while other threads are reading it, the outcome could be unpredicted or even lead to a program crash. In addition, applications usually rely on the static variable for security checks or maintaining a global state, which means that mutating the immutable data can also help attackers bypass security checks. \looseness=-1

\begin{lstlisting}[
language=rust, 
style=lst,
caption=A mismatched scope bug in \texttt{rgb} that allows viewing and modifying data of any type wrapped in \texttt{ComponentSlice<T>} as bytes (RUSTSEC-2020-0029\cite{rs-2020-0029}).,
label=listing:motivation-type-3,
mathescape=false]
pub trait ComponentBytes<T: Copy+Send+Sync+'static>
  where Self: ComponentSlice<T> {
  fn as_bytes_mut(&mut self) -> &mut [u8] {
    let slice = self.as_mut_slice();
    unsafe {
      slice::from_raw_parts_mut(
        slice.as_mut_ptr() as *mut _, ..)
    }
  }
}

impl<T> ComponentSlice<T> for [RGB<T>] { .. }

impl<T> RGB<T> {
  pub const fn new(r: T, g: T, b: T) -> Self {
    Self {b, g, r}
  }
}

// exploit for type III bug
let component: &'static str = "Hello, World!";
let new_rgb = RGB::new(component, .., ..);
let mut rgb_arr = [new_rgb; 3];
let bytes = rgb_arr.as_bytes_mut();
bytes[0] += component.len() as u8;
// now, we can modify static memory
println!("{}", rgb_arr[0]);
\end{lstlisting}

\subsection{Challenges and Insights}
\label{sec:challenge}
The Rust type system's features, such as ownership, trait bounds, and generic types, present challenges that cannot be directly addressed using existing methods in C and C++. We use the example of the \bthree bug to illustrate the three challenges.

\vspace{0.05in}
\noindent{\bf Interprocedural Type Conversion.} \label{motive:call-graph-construction}
{In Listing \ref{listing:motivation-type-3}, \TN identifies that the \texttt{as\_bytes\_mut} function performs a risky type conversion from a generic type to \texttt{u8} on line 7, potentially leading to a type confusion bug. To confirm the presence of this bug, \TN must determine if the type is converted between functions. Line 3 indicates that type \texttt{self} is initialized by a constructor function then passed to the current function. However, finding the constructor is a challenging problem that traditional call graphs cannot address. For instance, the type of \texttt{[RGB<T>]} implements \texttt{ComponentSlice} (refer to line 12) but must be initialized via the \texttt{new} function shown on line 15. Traditional call graphs can locate callers of \texttt{as\_bytes\_mut}, but the type constructor is not typically a direct caller (see lines 22 and 24). To address this issue, we identify the constructor by matching converted types to the return types of external functions via a new data structure called Property Graph}.

\vspace{0.05in}
\noindent{\bf Generic Type Resolution.} \label{motiv:generic-analysis}
{To predict the concrete types that can initialize generic type \texttt{T}, we first analyze trait bounds. The generic type \texttt{T}} is constrained by four trait bounds: \texttt{Copy}, \texttt{Send}, \texttt{Sync}, and \texttt{'static}.
Suppose we only enumerate the types directly involved in these trait bounds; we only get some internal types, such as \texttt{RGB<ComponentType>} and \texttt{BGR<ComponentType>}.
Since \texttt{ComponentBytes} is a public trait, the user can initialize it with external user-defined types, such as \texttt{str} for the generic type (line 21). 
To address such an issue, we extend {\it \analysisone} to {resolve the trait bounds}.
First, if these traits are also bounded by other traits, we need to parse the dependencies recursively.
Second, if the function is public,
we must consider all the primitive types and composite types that external users could
initialize. Therefore,
we collect traits and primitive types defined in standard libraries. {For the composite types, we build the \texttt{struct} type from primitive fields}. Finally, {\bugdetector can leverage the trait bounds to generate a type set. The type set includes all possible types that may be implicitly implemented and satisfy the trait bounds.}

\vspace{0.05in}
\noindent{\bf Alias Analysis.} \label{motiv:pointer-analysis}
{To verify the existence of the bug, it is crucial for us to determine whether the pointer alias remains valid after the type conversion (whether we can access \texttt{rgb\_arr} after line 24)}. 
The pointer type conversion on line 7 is translated to \texttt{\_8 = move \_9 as *mut u8 (PtrToPtr)} in the Mid-level Intermediate Representation (MIR \cite{MIR}) of the Rust compiler.
\texttt{move} represents the transfer of ownership of a value to another, which means \texttt{\_9} is not accessible anymore.
However, the previous alias of \texttt{\_9} still point to the same memory address.
Therefore, \texttt{slice} on line 4 is still accessible after the ownership of its mutable pointer (\texttt{\_9}) is transferred, leading to \bthree bug when accessing \texttt{rgb\_arr[0]} (line 27).
To precisely identify the alias relationship between the pointers, we analyze whether the pointer's ownership is transferred based on different forms of instruction in the Rust program.
In the Listing \ref{listing:motivation-type-3}, \analysistwo helps us verify that the parameter (\texttt{\&mut self}) points to the same memory location as the \texttt{u8} pointer (\texttt{\&mut [u8]}), and whether the parameter remains accessible after returning.
\looseness=-1



\subsection{Detection Scope}

We target the \bug arising from the pointer type conversions and specify the type conversion behavior to be implemented with \texttt{as} and \texttt{transmute} operations, representing the most fundamental ways to conduct type conversion in Rust. 
In particular,
\TN focuses on the three most prevalent bug types mentioned in~\autoref{sec:ty-bug}.
We do not consider the type conversion performed on non-pointer types, so the bugs such as integer overflow~\cite{rs-2024-0338, rs-2024-0016} arising from downcast are excluded.
Errors arising from foreign function interfaces~\cite{FFI} are also out of scope, as addressing them would require developing a system that is compatible with other programming languages' compilers.

\begin{figure}[t]
\centering
\includegraphics[width=\linewidth]{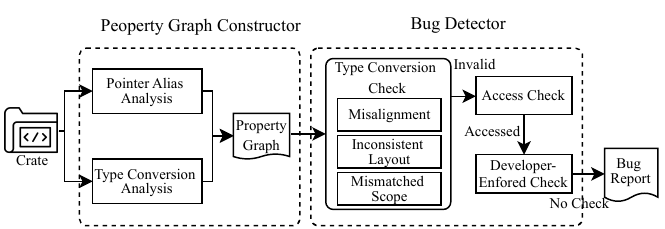}
\vspace{-10pt}
\caption{\label{fig:architecture} An overview of \TN.}
\label{architecture}
\end{figure}

\section{\TN}
\label{sec:design}

To facilitate the detection of \bugs, we design and implement a static analysis tool called \TN. 
This tool consists of two main components: \Tyanalyzer and \Bugdetector (see \autoref{architecture}). 
{
Given the Rust code, \tyanalyzer first utilizes the compiler to translate the source code into the MIR. It then performs \analysisone and \analysistwo to construct \pcg. The \pcg includes  the type conversion pairs and  the pointer alias graph. Besides, trait bounds are provided to \bugdetector for generic type resolution.}
Given the type conversion pairs, \bugdetector first performs the type conversion check with three different patterns to capture the invalid type pointer,
Then, the access check is run to determine whether the invalid type pointer can be accessed through the alias graph.
{
Finally, \bugdetector verifies if there are any developer-enforced checks implemented to handle the invalid type pointer.
If no, \TN produces a bug report with the problematic type conversion highlighted}. 

\subsection{Property Graph Constructor}\label{compone}
{
The goal of \tyanalyzer is to collect the required information and then integrate them into the \pcg, which can accelerate the interprocedural analysis of \bugdetector.
In addition to the results of \analysisone (\autoref{tyconvanalysis}) and \analysistwo (\autoref{aliasanalysis}),
each function is associated with its return types, assisting in finding the type constructor functions.
For example, 
given a function \texttt{new\_as\_slice},
it calls the type constructor function \texttt{new} and passes the constructed type to the function \texttt{as\_slice} as \texttt{src\_ty}.
To obtain \texttt{new} function, we match \texttt{src\_ty} with the return types of other functions in our \pcg, which could find all potential type constructors.
After that, we can further analyze the type conversion across the functions.
In addition to return types, the function including \texttt{unsafe} code will be marked for analysis in \bugdetector.
}

\begin{algorithm}[t]
\caption{\texttt{get\_trait\_bounds()}}\label{genanalysis}
\footnotesize
\SetKwFunction{callerbounds}{caller\_bounds}
\SetKwFunction{gettraitbnd}{get\_bnd\_by\_sig}
\SetKwFunction{gettypebnd}{get\_type\_bnd}
\SetKwFunction{getlocalimpls}{get\_local\_impls}
\SetKwFunction{selfty}{self\_ty}
\SetKwFunction{insert}{insert}
\SetKwFunction{intersect}{intersect}
\SetKwFunction{new}{new}
\SetKwFunction{hassupertrait}{has\_supertraits}
\SetKwFunction{isvisible}{is\_visible}
\SetKwInOut{Input}{Input}
\SetKwInOut{Output}{Output}

\SetAlgoLined

\Input{$fn$, $trait\_map$, $visible$}
\Output{$trait\_bounds$}

$trait\_bounds$ $\leftarrow$ HashSet::\new{};

\ForEach{$trait\_bnd \in fn.\gettraitbnd{}$}{
    \If{$trait\_bnd \in trait\_map$}{
        $trait\_bounds$.\insert{$trait\_bnd$};
    }
    \Else {
        \If{$trait\_bnd.\hassupertrait{} \&\&  visible$}{
            \tcp{call get\_trait\_bounds() on supertraits again}
        }
        \Else{
            $trait\_bounds$.\insert{$trait\_bnd$};
        }
    }
}

return $trait\_bounds$;
\end{algorithm}

\subsubsection{\Analysisone} \label{tyconvanalysis}
The \analysisone has three steps,
{namely, analyzing if type conversion includes any generic type, determining if generic type is converted across functions, and resolving the dependencies to collect the trait bounds on generic types}.
In the first step, \tyanalyzer directly keeps the concrete type pair if no generic type is included.
It starts with visiting the MIR's statements \cite{statement} and finding the ones of type conversion.
Rust's MIR is a simplified version of the Abstract Syntax Tree (AST) used for optimization.
It consists of statements and terminators \cite{terminator}. Statements represent intermediate operations such as assignments and variable initialization, and terminators define control flow decisions such as conditions or function calls.
In the statement of type conversion (\texttt{src\_ty}, \texttt{dst\_ty}), 
if both \texttt{src\_ty} and \texttt{dst\_ty} are the concrete types, \tyanalyzer will keep the type pair directly.
If one of the (\texttt{src\_ty}, \texttt{dst\_ty}) is a generic type, \tyanalyzer will move on to the second step, which is visibility analysis.

\vspace{0.05in}
\noindent{\bf Visibility Analysis.}
{
The visibility of a function decides how external users can call the function.
In Rust, functions and methods are both blocks of reusable code.
The difference between a function and a method is that the method is associated with a particular type or defined within a trait.
It is typically called using the \texttt{"."} operator on the type instance.
If the generic type conversion occurs in the method, we should determine whether the associated types can only be initialized by type constructor functions.
In such cases, we analyze the visibility of all associated types and recursively traverse the types fields if it is a \texttt{struct} type.
The visibility result represents if the type can be initialized by external users or limited by constructor functions.
If visible, trait bound analysis will be conducted to collect the type constraints for the generic type.
}
\looseness=-1

\vspace{0.05in}
\noindent{\bf Trait Bound Analysis.} 
{
We collect a set of traits from standard libraries, which are implemented by all primitive types in Rust, indicating the potential concrete types to replace generic types (see~\autoref{genanalysis}).
We also extract certain traits from external libraries used to prevent type confusion bugs, helping to reduce false alarms in bug detection.
For example, the trait \texttt{plain}\cite{plain} is always used to ensure that the memory layout is stable and initialized.
As we have confirmed the specific types implementing these traits (\texttt{trait\_map}), we utilize them as the endpoint of the traversal, effectively tackling the issue of implicit dependencies.
For each trait bound, \tyanalyzer first checks whether the trait is defined in the trait set.
If not defined, \tyanalyzer then checks whether the trait has dependencies (\texttt{has\_supertraits()}).
The output of this step also generates the type conversion pairs including generic type and associated with the trait bounds.
}

\subsubsection{Pointer Alias Analysis} \label{aliasanalysis}
\Analysistwo is used to construct an alias graph, which helps us determine the relation between pointers and how the pointer can be accessed (see \autoref{aliasgraph}).
The analysis is performed in MIR for semantic information~\cite{MIR}, e.g., whether a value is moved or borrowed and if the value is dead.
The nodes in the alias graph are collected from the \texttt{Local} in the MIR, which refers to the "variables and temporary values in the scope of function"~\cite{Local}. 
The edges between the nodes are updated when the MIR statement is in the forms of \texttt{StorageDead} and \texttt{Assign} form, where \texttt{Rvalue} is assigned to \texttt{Lvalue}. 
Based on the kinds of \texttt{Rvalue} appearing in the statement of \texttt{Assign}, pointers have different alias relationships.

\begin{equation} \label{assign}
    \begin{aligned}
        \text{a} &= \text{Ref(b)} \\
                     &= \text{RawPtr(b)} \\
                     &= \text{Cast::(PtrToPtr, Operand(b))} \\
                     &= \text{Cast::(Transmute, Operand(b))}
    \end{aligned}
\end{equation}

In \autoref{assign},
when the kind of \texttt{Rvalue} is \texttt{Ref} or \texttt{RawPtr}, which means a new reference or raw pointer \texttt{a} is created and points to the same memory location as \texttt{b}.
If the kind of \texttt{Rvalue} is \texttt{Cast}, especially on the pointers, \texttt{a} also points to the same location as \texttt{b}.
In our alias graph, we will create the edge from \texttt{a} to \texttt{b} to represent the alias relationship, where they are both \texttt{local}.
However, we disconnect the edge from \texttt{a} to \texttt{b} when the kind of \texttt{Operand} in \texttt{Cast} is \texttt{Move}.
The operand of \texttt{Move} means that the ownership of \texttt{b} is transferred to \texttt{a}, and \texttt{b} will no longer be accessible, so we disconnect the edge.
In addition to the statement in the \texttt{Assign} form, we also disconnect the edge in the form of \texttt{StorageDead}. 
Given \texttt{StorageDead(a)}, it is used to mark that the ownership of \texttt{a} is transferred and all pointers of \texttt{a} become invalid.
Therefore, we delete all edges created from \texttt{a} in our alias graph.
\begin{equation} \label{function}
    \text{a} = \text{Call(Fn, args<Operand(b)>, ..)}
\end{equation}

\begin{algorithm}[t]
\caption{\texttt{get\_alias\_graph()}}\label{aliasgraph}
\footnotesize
\SetKwFunction{Call}{Call}
\SetKwFunction{id}{id}
\SetKwFunction{push}{push}
\SetKwFunction{isderef}{is\_deref}
\SetKwFunction{lplace}{\lplace}
\SetKwFunction{newst}{new\_statement}
\SetKwFunction{newtm}{new\_terminator}
\SetKwFunction{newtg}{new\_taint\_graph}
\SetKwFunction{as}{As}
\SetKwFunction{tm}{Transmute}
\SetKwFunction{containsunsafe}{contains\_unsafe}
\SetKwFunction{declaredsafe}{declared\_as\_safe}
\SetKwFunction{getop}{get\_operand}
\SetKwFunction{getHIR}{get\_HIR}
\SetKwFunction{getMIR}{get\_MIR}
\SetKwInOut{Input}{Input}
\SetKwInOut{Output}{Output}
\SetAlgoLined

\Input{$fn$}
\Output{$alias\_graph$}
\ForEach{$st \in fn.statements$}{
    \If{$st \in Assign$}{
        ($lval$, $rval$) $\leftarrow$ ($st.lvalue()$, $st.rvalue()$);

        $op$ $\leftarrow$ $rval$.\getop{};
        
        $kind$ $\leftarrow$ $st.rvalue().kind()$;
        
        \If{$kind \in Ref | RawPtr | Cast::PtrToPtr | Transmute$}{
            // insert $rval$.\id{} to alias\_graph[$lval$.\id{}]
            
            \If{$op == Move$}{
                // delete $rval$.\id{} from alias\_graph[$lval$.\id{}]
            }   
        }
    }
    \ElseIf{$st \in StorageDead$}{
        $rval$ $\leftarrow$ $st.rvalue()$;
        
        // delete all elements from alias\_graph[$rval.\id{}$]
    }
}
\ForEach{$tm \in fn.terminators$}{
    $kind$ $\leftarrow$ $tm.kind()$;
    
    \If{$kind == \Call{func, args, dest}$}{
        \ForEach{$arg \in args$}{
            // insert arg.\id{} to alias\_graph[dest.\id{}];

            $op$ $\leftarrow$ $arg$.\getop{};

            \If{$op == Move$}{
                // delete $arg$.\id{} from alias\_graph[dest.\id{}]
            }
        }
    }
}

return $alias\_graph$
\end{algorithm}

{\autoref{function} presents a function call in MIR, where \texttt{args} works as a list of arguments that are passed to the function and \texttt{a} holds the return value}.
For each argument in \texttt{args}, we create the edge from \texttt{a} to \texttt{b}, but disconnect the edge if the operand on the argument is \texttt{Move}.
{In \bugdetector, the connection of \texttt{a} and \texttt{b} in the alias graph is leveraged to perform interprocedural alias analysis}.
, the \textit{alias\_graph} is constructed as a directed graph where the edge always starts from the \texttt{local} in \texttt{lvalue} to the one in \texttt{rvalue}.
inal{
When identifying pointer aliasing, we will check whether two nodes have common descendents in \textit{alias\_graph}.
Finding the common descendent represents that one alias of the \texttt{src\_ty}'s pointer is aliased with the \texttt{dst\_ty}'s pointer,
then \bugdetector will collect all descendants while traversing the graph with breadth-first search~\cite{Breadthf77online} from two nodes, then find whether there is an intersection between two sets of descendants.
}

\subsection{Bug Detector} \label{compthree}
{
Bug detector focuses on the marked functions in \pcg (with \texttt{unsafe})}, capturing \bugs in three steps. 
First, given the pairs of type sets generated by \analysisone, \checkone is performed to find the invalid type pointer following three kinds of bug patterns.
Second, \checktwo is used to find the alias of the invalid type pointer
based on \analysistwo.
Based on the alias graph, it checks if the pointer is accessed in the function or accessible to the caller function. 
Third, verifying the absence of \tc helps reduce the false alarms of bugs.
{All three steps are combined with interprocedural analysis based on \pcg}.

\begin{table*}[t]
\centering
\caption{\Checkone{s}.}
\begin{threeparttable}
\footnotesize
\begin{tabular}{c l l l} 
 \toprule
 Bug Type & Con $\rightarrow$ Con$^\ddagger$ & Con $\rightarrow$ Gen  & Gen $\rightarrow$ Con \\ 
 \midrule
 {Type I} &
 
 \thead[l]{
 \textbf{Input}: \textit{src\_ty}, \textit{dst\_ty} \\ 
 \textbf{If} \textit{src\_ty}.align \% \textit{dst\_ty}.align != 0 \\ 
 ~ ~ ~ ~ mark
 } &
 
 \thead[l]{
 \textbf{Input}: \textit{src\_ty}, \textit{ty\_set} \\
 \textbf{If} \textit{ty\_set}.is\_empty() \\ 
 ~ ~ ~ ~ mark \\
 \textbf{Else} \\
 ~ ~ ~ ~ replace \textit{dst\_ty} with each in \textit{ty\_set} \\
 ~ ~ ~ ~ run again in (Con $\rightarrow$ Con)} &

  \thead[l]{
 \textbf{Input}: \textit{dst\_ty}, \textit{ty\_set} \\
 \textbf{If} \textit{ty\_set}.is\_empty() \& \textit{dst\_ty}.align != 1 \\
 ~ ~ ~ ~ mark \\ 
 \textbf{Else If} ty\_set.not\_empty() \\ 
 ~ ~ ~ ~ replace \textit{src\_ty} with each in ty\_set \\
 ~ ~ ~ ~ run again in (Con $\rightarrow$ Con)} \\
 \hline
 {Type II} &
 
 \thead[l]{
 \textbf{Input}: \textit{src\_ty}, \textit{dst\_ty} \\
 \textbf{If \textit{src\_ty}} $\rightarrow$ \textit{unstable\_ty} \\
 ~ ~ ~ ~ \textbf{If \textit{dst\_ty}} $\rightarrow$  (\textit{stable\_ty} | \textit{unstable\_ty}') \\
 ~ ~ ~ ~ ~ ~ ~ ~ mark
 } &
 
 \thead[l]{
 \textbf{Input}: \textit{src\_ty}, \textit{ty\_set} \\
 \textbf{If} \textit{ty\_set}.is\_empty() \\
 ~ ~ ~ ~ \textbf{If \textit{src\_ty}} $\rightarrow$ \textit{unstable\_ty} \\
 ~ ~ ~ ~ ~ ~ ~ ~ mark \\
 \textbf{Else} \\
 ~ ~ ~ ~ replace \textit{dst\_ty} with each in \textit{ty\_set} \\
 ~ ~ ~ ~ run again in (Con $\rightarrow$ Con)} &

  \thead[l]{
 \textbf{Input}: \textit{dst\_ty}, \textit{ty\_set} \\
 \textbf{If} \textit{ty\_set}.is\_empty() \\
 ~ ~ ~ ~ \textbf{If \textit{dst\_ty}} $\rightarrow$ (\textit{stable\_ty} | \textit{unstable\_ty}') \\
 ~ ~ ~ ~ ~ ~ ~ ~ mark \\
 \textbf{Else} \\
 ~ ~ ~ ~ replace \textit{src\_ty} with each in \textit{ty\_set} \\
 ~ ~ ~ ~ run again in (Con $\rightarrow$ Con)
 } \\
 \hline
 Type III &
 
 \thead[l]{
 \textbf{Input}: \textit{src\_ty}, \textit{dst\_ty} \\
 \textbf{if \textit{src\_ty}} $\rightarrow$ \textit{weak\_ty} \\
 ~ ~ ~ ~ \textbf{If \textit{dst\_ty}} $\rightarrow$ \textit{strict\_ty} \\
 ~ ~ ~ ~ ~ ~ ~ ~ mark \\
 \textbf{Else If \textit{src\_ty}} $\rightarrow$ \textit{strict\_ty} \\
 ~ ~ ~ ~ \textbf{If \textit{dst\_ty}} $\rightarrow$ \textit{mut weak\_ty}  \\
 ~ ~ ~ ~ ~ ~ ~ ~ mark } &
 
 \thead[l]{
 \textbf{Input}: \textit{src\_ty}, \textit{ty\_set} \\
 \textbf{If} \textit{ty\_set}.is\_empty() \\
 ~ ~ ~ ~ \textbf{If \textit{src\_ty}} $\rightarrow$ \textit{weak\_ty} \\
 ~ ~ ~ ~ ~ ~ ~ ~ mark \\
 \textbf{Else If \textit{src\_ty}} $\rightarrow$ \textit{strict\_ty}
  \&\& \textit{mut dst\_ty} \\
 ~ ~ ~ ~ ~ ~ ~ ~ mark \\
 \textbf{Else} \\
 ~ ~ ~ ~ replace \textit{dst\_ty} with each in ty\_set \\
 ~ ~ ~ ~ \textbf{If (s,d)$^\star$} $\rightarrow$ (\textit{weak\_ty}, \textit{strict\_ty}) \\
 ~ ~ ~ ~ ~ ~ ~ ~ mark \\
 ~ ~ ~ ~ \textbf{Else If (s,d)} $\rightarrow$ (\textit{strict\_ty}, \textit{mut weak\_ty}) \\
 ~ ~ ~ ~ ~ ~ ~ ~ mark } &
 
 \thead[l]{
 \textbf{Input}: \textit{dst\_ty}, \textit{ty\_set} \\
 \textbf{If} \textit{ty\_set}.is\_empty() \\
 ~ ~ ~ ~ \textbf{If \textit{dst\_ty}} $\rightarrow$ \textit{strict\_ty} \\
 ~ ~ ~ ~ ~ ~ ~ ~ mark \\
 ~ ~ ~ ~ \textbf{Else If \textit{dst\_ty}} $\rightarrow$ \textit{weak\_ty} \&\& \textit{mut dst\_ty} \\
 ~ ~ ~ ~ ~ ~ ~ ~ mark \\
 \textbf{Else} \\
 ~ ~ ~ ~ replace \textit{src\_ty} with each in \textit{ty\_set} \\
 ~ ~ ~ ~ \textbf{If (s,d)} $\rightarrow$ (\textit{weak\_ty}, \textit{strict\_ty}) \\
 ~ ~ ~ ~ ~ ~ ~ ~ mark \\
 ~ ~ ~ ~ \textbf{Else If (s,d)} $\rightarrow$ (\textit{strict\_ty}, \textit{mut weak\_ty}) \\
 ~ ~ ~ ~ ~ ~ ~ ~ mark } \\

 \bottomrule
\end{tabular}
\begin{tablenotes}
    \footnotesize
    \item $^\ddagger$ Con: concrete type; Gen: generic type; $^\star$ (s,d): (\textit{src\_ty}, \textit{dst\_ty}).
\end{tablenotes}
\end{threeparttable}
\label{bugdetectors}
\end{table*}

\subsubsection{Type Conversion Check}
We categorize the type conversion (\textit{src\_ty}, \textit{dst\_ty})into three possible scenarios: (Con $\rightarrow$ Con), (Con $\rightarrow$ Gen), and (Gen $\rightarrow$ Con).
The conversion between (Gen $\rightarrow$ Gen) is excluded since we observe that such a conversion would be rejected by the \texttt{TypeId} check\cite{TypeIdin2:online}. The check strictly requires two types sharing the same layout.
{When generic type conversion errors are identified, the trait bounds linked to the generic type are  mapped to the \textit{ty\_set}, which has been verified to implement these traits in \tyanalyzer.}
The detection logic for each bug type and each scenario is shown in \autoref{bugdetectors}.

\vspace{0.05in}
\noindent{\bf \Bdone (Type I).}
Misalignment detector can easily compute the alignments to identify the bugs; however, it is not possible to predict the alignment of generic types that depend on runtime input. To solve the challenge, we will use \textit{ty\_set} to simulate the input to generic types.

\noindent\emph{Bug Definition.}
When \textit{src\_ty}'s alignment is not a multiple of \textit{dst\_ty}'s alignment, it will create a misaligned pointer. 

\noindent\emph{Type Conversion.}
In the scenario of (Con $\rightarrow$ Con), \bdone directly locates the type conversion by computing the violation of alignment requirements (i.e., \texttt{\textit{src\_ty}.align \% \textit{dst\_ty}.align != 0}), and then we will mark the \textit{dst\_ty}'s pointer as an invalid type pointer.
In the scenario of (Con $\rightarrow$ Gen), we need to traverse all candidate types in \textit{ty\_set} to ensure each type obeys the alignment requirements. If any candidate type violates the requirement, we mark it as an invalid pointer. When \textit{ty\_set} is empty, it means that the generic type can be initialized with arbitrary types since no trait bounds are found. In this case, we will also mark \textit{dst\_ty}'s pointer as an invalid pointer.
In the scenario of (Gen $\rightarrow$ Con), the detector follows the same logic as in (Con $\rightarrow$ Gen) to mark the invalid pointer. The difference is that \textit{dst\_ty} can not be aligned to only one byte even when the \textit{ty\_set} is empty. The reason is that any memory address can be a multiple of one where the misaligned pointer will not be created.
In some cases, \bdone may fail since the types are imported from external packages. To solve the challenge, we heuristically extract the information from the symbol names of types based on the cases we have studied (e.g., extract \texttt{u8} from \texttt{external::u8\_bytes}).
Since we only run our tool on the machine of 64-bit architecture; however, some type's alignment is platform-specific, where the value changes based on different architectures.
Take \texttt{usize} and \texttt{isize} for example, on a 32-bit target, they are aligned to 4 bytes while on a 64-bit target, they are aligned to 8 bytes.
In \Bdone, we will consider different alignment values for these types in the type conversion.

\vspace{0.05in}
\noindent{\bf \Bdtwo  (Type II).}
To detect the inconsistent layout bug, we define two type sets: \textit{unstable\_ty} and \textit{stable\_ty}. 
\textit{unstable\_ty} represents the type that can change the memory layout at runtime (e.g., \texttt{struct}, \texttt{union}, trait object), where the compiler preserves the rights to insert padding bytes or reorder the fields. Another type set \textit{stable\_ty} consists of scalar types (e.g., bool, char, integers). \Bdtwo will perform further analysis on the representation of types (e.g., \texttt{repr(Rust)}, \texttt{repr(transparent)}, \texttt{repr(C)}). 
Any type conversion in \textit{unstable\_ty} set or across \textit{unstable\_ty} and \textit{stable\_ty} sets would be recognized as a problematic type conversion and create an invalid type pointer. In addition, we need to combine with \textit{ty\_set} to extend the scenarios of generic type conversion.

\noindent\emph{Bug Definition.}
When the layout of \textit{src\_ty} is not stable and inconsistent to \textit{dst\_ty}, it will create an invalid type pointer. 

\noindent\emph{Type Conversion.} 
If the detector finds the conversion happens in (\textit{unstable\_ty} $\rightarrow$ \textit{stable\_ty}), it will mark the \textit{dst\_ty}'s pointer as an invalid type pointer since the padding bytes can be exposed when we accessed them as a scalar type.
The second pattern is (\textit{unstable\_ty} $\rightarrow$ \textit{unstable\_ty}), \bdtwo will further check if they follow the same Application Binary Interface (ABI)~\cite{unsoundcollectiontransmute}, which determines if \textit{src\_ty} and \textit{dst\_ty} share same layout based on their symbol name of type. If they have different type symbol names, \textit{dst\_ty}'s pointer will also be marked as an invalid type pointer.
In the scenarios of (Con $\rightarrow$ Gen) and (Gen $\rightarrow$ Con), they follow the same logic to check when the arbitrary types or the limited types in \textit{ty\_set} could make the type conversion match the two patterns above. 

\vspace{0.05in}
\noindent{\bf \Bdthree (Type III).}
In order to find the bug efficiently, we define two type sets based on the scope of values: \textit{weak\_ty} and \textit{strict\_ty}.
\textit{weak\_ty} represents the type that has a weak constraint on its bit pattern, such as integer and float type. In contrast, \textit{strict\_ty} means a strong limitation on the bit pattern, such as bool, string, and character.
If the type is found to be a composite type, which has multiple fields, \bdthree will analyze each field and define it as \textit{strict\_ty} if one of the fields is included in \textit{strict\_ty}.
The type conversion between \textit{weak\_ty} and \textit{strict\_ty} can create a type with an invalid bit pattern.

\noindent\emph{Bug Definition.} 
There are two patterns of conversion that can create an invalid type: 
1) \textit{src\_ty} belongs to \textit{weak\_ty} while \textit{dst\_ty} belongs to \textit{strict\_ty}. 
2) \textit{src\_ty} and \textit{dst\_ty} are in \textit{strict\_ty} and \textit{weak\_ty}, while \textit{dst\_ty}'s pointer is mutable.
In these two types of conversions, an invalid type pointer can be created.

\noindent\emph{Type Conversion.} 
If the detector finds the conversion in (\textit{weak\_ty} $\rightarrow$ \textit{strict\_ty}), it will mark the \textit{dst\_ty}'s pointer as an invalid type pointer since the bit pattern of \textit{src\_ty} could be invalid for \textit{dst\_ty}.
When the conversion is found in (\textit{strict\_ty} $\rightarrow$ \textit{weak\_ty}), \bdthree will take a further analysis on whether the \textit{dst\_ty} is mutable since changing the bit pattern of \textit{dst\_ty} can also make \textit{src\_ty} invalid.
In the scenarios of (Con $\rightarrow$ Gen) and (Gen $\rightarrow$ Con), the detector follows the same logic
to check whether the type conversion is performed between \textit{weak\_ty} and \textit{strict\_ty} with mutability analysis.
\looseness=-1

\subsubsection{Access Check}
\label{accesscheck}

\Checktwo is performed to analyze how the invalid type pointer captured by \Checkone can be accessed.
The analysis can be separated into two steps:
1) check whether the pointer is accessed in the function,
and 2) analyze whether the pointer is accessible for the caller function.
As the first step, to check whether \textit{dst\_ty}'s pointer is accessed in the function, we focus on the dereference in statements and the unsafe function calls in the terminators of the MIR.
%
For statements, we check whether \textit{dst\_ty}'s pointer is aliased with the dereferenced pointer.
For unsafe function calls, we collect a list of unsafe functions that are widely used in the core libraries of Rust, such as \texttt{ptr::read/copy}, \texttt{ptr::as\_ref}, and \texttt{slice::from\_raw\_parts}, {which requires the pointer refers to an aligned, consecutively initialized type.
The access list for \bdthree also includes other APIs such as \texttt{str::from\_utf8\_unchecked} or \texttt{CStr::from\_ptr}.
These functions require types to be encoded with the specific bit patterns}.
\Checktwo will verify whether the pointer passed in these unsafe functions is aliased with the \textit{dst\_ty}'s pointer.
\looseness=-1

In the second step, to check whether \textit{dst\_ty}'s pointer is accessible for the caller function, we analyze whether the \textit{dst\_ty}'s is aliased with the return type only when the return type is a reference.
When the return type is a raw pointer, accessing it requires the \texttt{unsafe} block since Rust does not guarantee the safety of the raw pointer.
Since using \texttt{unsafe} highlights the responsibility for the bugs and we only consider the function that performs the problematic type conversion to be the culprit of bugs, we will set up the requirement for the return type to be a reference in the second step.
%
%
%
After \checktwo ensures that the pointer of invalid \textit{dst\_ty} can be accessed, the bug report will be generated as the output of \bugdetector.

\subsubsection{\TC Analysis}
\label{sec:manualtypecheck}
{\TC is usually used by the developers to prevent type confusion bugs manually. Through examining these checks, we can confirm that the developer has handled the type conversion errors, further reducing the false alarms of \TN{}.
We categorize them into two scenarios: \textit{Pre Type Check} and \textit{Post Type Check}, where the check inserted before and after type conversions, respectively.}

{We summarize various patterns of developer-enforced checks that address type confusion bugs individually.
First, the pre type checks used to prevent misalignment bugs include calling \texttt{align\_of} and \texttt{alloc}, which are used to check and assign memory layout before the type conversion. 
There is also a post type check that the developer uses to safely load the misaligned type e.g., \texttt{read\_unaligned}. 
Second, for the inconsistent layout bug, pre type check is used to guarantee the memory is completely initialized. 
The typical patterns include using \texttt{size\_of} to restrict the size at run-time.
For example, if the struct contains two fields of \texttt{u32} types, developers can check if the size of the struct type is 8 bytes, further ensuring that no padding bytes are inserted.
The post type checks such as \texttt{ptr::write}, which can be used to access the uninitialized memory, will also be detected before \TN raises the alarms for the bugs.
Detecting developer-enforced checks for scope mismatch bugs is challenging because developers often use runtime value comparisons.}

\subsubsection{Integration of Interprocedural Analysis}
{
Interprocedural analysis plays an essential role in confirming the presence of the type confusion bug. In this section, we describe how it is incorporated into \bugdetector.}

\vspace{0.05in}
\noindent{\bf Type Conversion Check.}
{We can use it to identify type conversions between functions. For instance, consider (Con $\rightarrow$ Con), where we notice an unsafe type casting from a \texttt{u8} pointer to a \texttt{u16}. According to our misalignment bug criteria, this should trigger an alert. Nevertheless, through interprocedural analysis, we identify a type conversion from a \texttt{u16} pointer back to \texttt{u8} in the caller function. As a result, the \texttt{src\_ty} should be \texttt{u16} and properly aligned to two bytes, which means that there is no any misalignment. Considering another case with generic type (Gen $\rightarrow$ Con) and type conversion being detected in a method, we leverage \pcg to identify the type constructor function and analyze the type conversion pairs in the method.}

\vspace{0.05in}
\noindent{\bf Access Check.}
{Given that the type may not be accessible within the current function, we also examine the callee functions, such as a raw pointer dereference. Additionally, we gather certain \texttt{unsafe} standard library functions, which involve type access, accelerating the verification of type access.}

\vspace{0.05in}
\noindent{\bf \TC Analysis.}
{\TC could also exist in external functions. In other words, it can be implemented in callers, type constructors, or callees. Thus, \TN must analyze all reachable functions to locate the related type checks.}

\subsection{Implementation} \label{impl}
\TN is developed with 5249 lines of code in Rust, utilizing \textit{rustc} and fully integrating with \textit{Cargo}, Rust's official package manager. \TN focuses on target files that can be compiled into an executable or a library \cite{cargotarget54online}. Using \textit{Cargo}, we address dependency issues prior to compilation and identify all targets in the package suitable for analysis. Compilation of these target files is done through \textit{rustc}. Upon completion, \TN is activated within the \texttt{after\_analysis} callback function of the \textit{rustc} driver, which is triggered by \textit{rustc} following the generation of Rust compiler's MIR, allowing us to employ the resulting MIR data as input for \tyanalyzer to start the analysis.

The workflow of \TN{} can be divided into two phases: 1) detecting if the type conversion generates a problematic \texttt{dst\_ty}, and 2) checking if the problematic type is accessed. With pairs of type sets generated by \analysisone, we can find a problematic type conversion even when a generic type is involved. With the alias graph built by \analysistwo, we can track how the pointer can be accessed. The type conversion pairs and alias graph are stored in \pcg, which accelerates the interprocedural analysis to obtain the information of external functions. For interprocedural analysis, we introduced a depth limitation to avoid the path explosion problem. We set the path length to 1 (tracing only the immediate caller or callee function), aligning with that in Rupta~\cite{rupta}.

\begin{table}[t]
\centering
\caption{
{Bugs identified by \TN; note that we have only listed the bugs that have been reported for over three months as of January 1, 2025.}
}
\resizebox{\columnwidth}{!}
{
\begin{threeparttable}
\renewcommand\arraystretch{1.1}
\setlength{\columnsep}{1pt}
\begin{tabular}{l r r l c c} 
 \toprule
 Package & Version & Stars & Bug Types$^\ddagger$ and Numbers & Status$^*$ & Patched$^\dagger$ \\ 
 \midrule
 candle-core & 0.4.1 & 13.2k & Con $\rightarrow$ Gen: I:3 (\texttt{as}) & \CIRCLE & - \\
 \hline
 py-spy & 0.3.14 & 11k & Con $\rightarrow$ Con: I:1 (\texttt{as}) & \Circle & - \\
 \hline
 fyrox-core & 0.27.0 & 7.1k & \parbox{5cm}{Con $\rightarrow$ Gen: I:1 (\texttt{as}) \\ Gen $\rightarrow$ Con: II:4 (\texttt{as}), III:2 (\texttt{as})} & \CIRCLE & \checkmark \\
 \hline
 gfx-backend-gl & 0.9.0 & 5.2k & Con $\rightarrow$ Gen: I:1 (\texttt{as}) & \CIRCLE & - \\
 \hline
 silicon & 0.5.2 & 3.1k & Con $\rightarrow$ Con: II:1 (\texttt{as}) & \Circle & - \\
 \hline
 webrender & 0.61.0 & 3k & Con $\rightarrow$ Con: I:2 (\texttt{as}) & \CIRCLE & - \\
 \hline
 spl-token-swap & 3.0.0 & 2.3k & Con $\rightarrow$ Gen: I:1 (\texttt{as}), III:1 (\texttt{as}) & \CIRCLE & - \\
 \hline
 scryer-prolog & 0.9.4 & 1.9k & Con $\rightarrow$ Con: I:6 (\texttt{transmute}) & \CIRCLE & - \\
 \hline
 libafl & 0.10.1 & 1.6k & Con $\rightarrow$ Con: I:3 (\texttt{as}), III:4 (\texttt{as}) & \CIRCLE & \checkmark \\
 \hline
 mesalink & 1.1.0 & 1.5k & Gen $\rightarrow$ Con: II:1 (\texttt{as}) & \Circle & - \\
 \hline
 fontdue & 0.8.0 & 1.2k & Gen $\rightarrow$ Con: I:1 (\texttt{transmute}) & \CIRCLE & - \\
 \hline
 pprof & 0.13.0 & 1k & \parbox{3cm}{Con $\rightarrow$ Con: II:1 (\texttt{as}) \\ Con $\rightarrow$ Gen: I:1 (\texttt{as})} & \CIRCLE & \checkmark \\
 \hline
 rendy-core & 0.5.1 & 814 & Gen $\rightarrow$ Con: II:2 (\texttt{as}) & \LEFTcircle & - \\
 \hline
 rendy-util & 0.4.1 & 814 & Gen $\rightarrow$ Con: II:2 (\texttt{as}) & \LEFTcircle & - \\
 \hline
 sciter-rs & 0.5.58 & 784 & Gen $\rightarrow$ Con: I:1 (\texttt{as}) & \CIRCLE & - \\
 \hline
 rosrust & 0.9.11 & 728 & Gen $\rightarrow$ Con: II:3 (\texttt{as}) III:1 (\texttt{as}) & \LEFTcircle & - \\
 \hline
 cortex-m & 0.7.7 & 669 & Gen $\rightarrow$ Con: II:1 (\texttt{as}) & \CIRCLE & \checkmark \\
 \hline
 rafx-base & 0.0.15 & 574 & Gen $\rightarrow$ Con: II:2 (\texttt{as}) & \LEFTcircle & - \\ 
 \hline
 xous & 0.9.50 & 500 & Con $\rightarrow$ Gen: I:2 (\texttt{as}), III:2 (\texttt{as}) & \CIRCLE & \checkmark \\
 \bottomrule
\end{tabular}
\begin{tablenotes}
    \item $^*$ Bug status: \Circle: Reported; \CIRCLE: Confirmed by Vendors; \LEFTcircle: Verified by PoC.
    \item $^\dagger$ Patch status: \checkmark: already patched; -: unpatched.
    \item $^\ddagger$ Bug type: Con: concrete type; Gen: generic type; [I, II, III]: bug type.
\end{tablenotes}
\end{threeparttable}
}
\label{top1k-zerodays}
\end{table}

\section{Evaluation}

\noindent{\bf Dataset Collection.} We gathered packages from \texttt{crates.io}, the Rust community's crate registry. To ensure comprehensive bug detection, The dataset consists of the packages ranked by download counts and the number of GitHub stars (as of September 1, 2024), and each of them has more than 500 stars. 
Regarding package size, the largest package contains 510k LoC, with an average package size of 9k LoC.

\vspace{0.05in}
\noindent{\bf Experiment Setup.} We built \TN{} and conducted experiments on a server with 48-core Intel Xeon CPU ES-2630 and 256 GB memory. The server was deployed with Ubuntu 22.04 and \texttt{rustc} 1.72.0-nightly. For each package, we set the preparation (dependencies resolution and compilation) time threshold to 20 minutes and \TN detection time threshold to 2 minutes. We ran \TN on the 3,000 packages for detection.

\subsection{Bug Detection Results} \label{evaluation:effectiveness}

\begin{figure}[t]
    \centering
    \includegraphics[width=\linewidth]{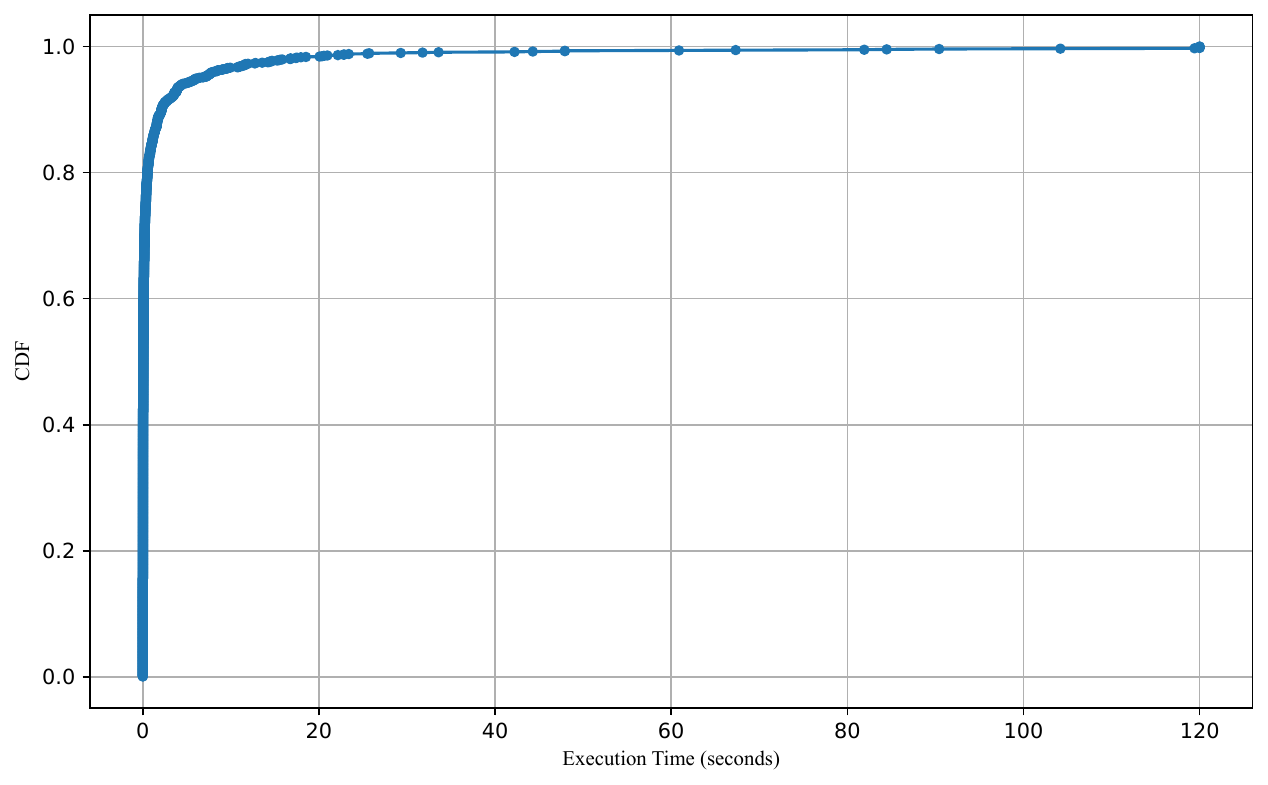}
    \caption{Detection time of \TN.}
    \label{fig:cdf-execution}
\end{figure}

First, we ran \TN on the RUSTSEC dataset of existing type confusion bugs 
{as of September 1, 2024 (see \autoref{rustsec-details})}.
\TN is capable of detecting all existing \bugs of three types. Second, {we ran \TN with 16-thread to scan the 3,000 packages.} {It ran 18 hours in total; the majority of time was spent on resolving dependencies and compilation. The detection time only took about 80 mins and 94 bugs were reported. For the 3,000 packages, 618 (20.6\%) failed due to compiler version, 37 (1.2\%) did not have proper cargo metadata, and 309 (10.3\%) failed with custom build options, which cannot be resolved automatically and required manual work. All the remaining 2,036 packages (67.9\%) could successfully compile within 20 minutes. Among them, 282 did not have \texttt{/bin} or \texttt{/lib} targets for detection, so TypePulse did not execute on them. For 1,754 packages that TypePulse ran detection on, the average detection time was 1.71s. Among the 1,754 packages, 1,483 were completed within 1s; 1,652 were completed within 5s; only 59 needed more than 10s to complete; 5 packages did not finish execution within 2 minutes. The CDF figure of the detection time is shown in~\autoref{fig:cdf-execution}.} {We manually verified these bug reports and confirmed that {71} of them were true positive, indicating an overall precision of {75.5\%}.} {The information on packages and bugs that have either been confirmed or reported for over three months is provided in \autoref{top1k-zerodays}}. Many bugs we detected are not trivial. {Among the 71 type confusion bugs detected by \TN, there are {50} bugs found in packages with more than 1,000 stars on GitHub}. These packages are well known in the Rust community and are usually maintained well by professional teams. For example, {\tt candle-core}~\cite{candle-core} is developed by a famous AI company (Hugging Face)~\cite{HuggingFace}. Therefore, our results suggest that even domain experts may write error-prone code in Rust.

\begin{table}[]
\centering
\caption{{The results of detectors on top 3,000 packages.}}
\footnotesize
\begin{threeparttable}
\begin{tabular}{lcrrrr}
\toprule
\multicolumn{1}{c}{\multirow{2}{*}{Detector}}                                & \multirow{2}{*}{Metrics} & \multicolumn{3}{c}{Bug Types}                                            & \multicolumn{1}{c}{\multirow{2}{*}{Overall}} \\ \cline{3-5}
\multicolumn{1}{c}{}                                                         &                          & \multicolumn{1}{c}{I} & \multicolumn{1}{c}{II} & \multicolumn{1}{c}{III} & \multicolumn{1}{c}{}                         \\ \hline
\multirow{3}{*}{TypePulse}                                                   & TP & {32} & {24} & {15} & {71}    \\
                                                                             & FP & {6}                     & {4}                      & {13}                       & {23}    \\
                                                                             & Precision                & {84.2\%}                     & {85.7\%}                      & {53.6\%}                       & {75.5\%}    \\ \hline
\multirow{3}{*}{\begin{tabular}[c]{@{}l@{}}TypePulse\\ w/o IPA\end{tabular}} & TP                       & {32}                     & {24}                      & {13}                       & {69}   \\
                                                                             & FP                       & {11}                     & {6}                      & {13}                       & {30}    \\
                                                                             & Precision                & {74.4\%}                     & {80\%}                      & {50\%}                       & {69.7\%}    \\
\bottomrule
\end{tabular}
\begin{tablenotes}
\item IPA: Interprocedural Analysis.
\end{tablenotes}
\end{threeparttable}
\label{detection-table}
\end{table}

\vspace{0.05in}
\noindent{\bf Capability to Find New Bugs.} 
To facilitate the resolution of these bugs, we also created a Proof of Concept (PoC) to report identified issues to package maintainers, which requires locating undefined behaviors in packages based on the diagnostic information produced by \TN. To trigger potential bugs, we need to create the appropriate \textit{src\_ty} and \textit{dst\_ty} before invoking the problematic conversion. We consider two distinct scenarios. First, when the source and destination variables are openly accessible, such as through public members within a class, we can directly create these variables. Second, when the source and destination variables are private and not directly accessible, {a constructor is required to initialize the variables. Additionally, \TN helps find feasible constructor functions, which also accelerates the PoC generation}.

At the time of writing, we have documented all the issues ({71}) that we have manually confirmed, and {32} of these have received acknowledgment responses from the package maintainers. After notifying the maintainers, we further reported these issues to the \rs advisories and CVE database. So far, we have received {six \rs IDs and one CVE ID}, and are awaiting confirmation and releases for the remaining ones. 
\looseness=-1

\subsection{False Positive Analysis} \label{evaluation:fp}

According to~\autoref{detection-table}, there are {23} instances of false positives. There are 2 arising from misidentifying function visibilities and {3} from developers' tricks to prevent invalid pointer exposure. Here we investigate the remaining {18} cases that arise from misinterpretation during the \tc analysis. It shows that \TN cannot understand complex condition semantics in checks, especially when the check is implemented in an unusual way. Considering Listing \ref{listing:xous} for example, \TN considers the method \texttt{to\_str} could construct an illegal string from an array of \texttt{u8}, which might contain a non-utf8 character. After manually studying the type constructor function (\texttt{new}) and the method used to insert characters into the string (\texttt{push}), we find that the characters in the string are restricted to be utf8 by the encoding at line 15 (\texttt{encode\_utf8}). This complex encoding makes it difficult for \TN to avoid false alarms in detecting Mismatched Scope Bugs. So far, \TN can only detect standard developer-enforced check patterns and several unsafe APIs with clear safety documentation. It is challenging for \TN to understand such encoding operations. We will extend the capability of \TN to interpret \tc in future work.

\begin{lstlisting}[
language=rust, 
style=lst,
caption=The false positive case of Mismatched Scope bug.,
label=listing:xous,
mathescape=false]
impl<const N: usize> String<N> {
  pub fn new() -> String<N> {
    String { bytes: [0; N], len: 0 }
  }
  pub fn to_str(&self) -> &str {
    unsafe { 
    str::from_utf8_unchecked(&self.bytes[0..]) }
  }
  pub fn push(&mut self, ch: char) -> Result<..> {
    match ch.len_utf8() {
      1 => { ... }
      _ => {
        let mut bytes: usize = 0;
        let mut data: [u8; 4] = [0; 4];
        let subslice = ch.encode_utf8(&mut data);
        // ...
\end{lstlisting}

\subsection{Impacts of Interprocedural Analysis} \label{interimpacts}

We conduct an ablation study to demonstrate \TN{}'s capabilities in reducing false positives by disabling interprocedural analysis (refer to the second row in~\autoref{detection-table}). \TN{} reduces 6 more false positives and detects 2 more true positives with interprocedural analysis enabled. Scanning 3k packages, \TN obtains the results with precision {75.5\%}, consisting of {71} True Positives and {23} False Positives. After disabling the interprocedural analysis, \TN's precision is reduced to only {69.7\%}, consisting of {69} True Positives and {30} False Positives. This comparison result highlights that interprocedural analysis can significantly enhance precision by analyzing the context across functions. First, it can detect the positive cases relying on external functions to decide the bit patterns -- 2 more mismatched scope bugs were detected. Second, it can reduce the false alarms in cases with developer-enforced checks -- {7 more False Positives were reduced}. For example, in the package of \texttt{arrow-buffer}, disabling interprocedural analysis will cause four more false positive cases of misalignment bugs. The Listing \ref{listing:arrow-buffer} shows a false positive case mitigated by interprocedural analysis. \TN first locates suspicious type conversion at line 20 since it finds that the pointer of \texttt{buffer} is cast to arbitrary generic type. However, the generic type cannot be controlled by attackers since \texttt{as\_slice} is a method relying on \texttt{BufferBuilder} and casting \texttt{MutableBuffer} to generic type. The constructor function of \texttt{MutableBuffer} cannot be located by traditional interprocedural analysis because it is not the caller of the method. \TN implements the functionality to find out the constructor functions by matching the type to the ones returned from other functions. In this example, \TN locates the function \texttt{with\_capacity}, which returns \texttt{MutableBuffer} as the constructor function. After analyzing the constructor function, \TN finds that the constructor function already guarantees the alignment of type with \texttt{Layout::from\_size\_align}; therefore, this should be a false alarm. Without interprocedural analysis, \TN cannot detect the developer-enforced check implemented in the constructor function.

\begin{lstlisting}[
language=rust, 
style=lst,
caption=The false positive case resolved by interprocedural analysis.,
label=listing:arrow-buffer,
mathescape=false]
impl MutableBuffer {
  #[inline]
  pub fn with_capacity(capacity: usize) -> Self {
    let layout = Layout::
      from_size_align(capacity, ALIGNMENT).unwrap();
    let data = match layout.size() {
      0 => dangling_ptr(),
      _ => {
        let raw_ptr = unsafe { 
          std::alloc::alloc(layout) 
        };
    // ...
    Self { data, len: 0, layout }
// ...
impl<T: ArrowNativeType> BufferBuilder<T> {
  #[inline]
  pub fn as_slice(&self) -> &[T] {
    // ...
    unsafe { std::slice::from_raw_parts(
      self.buffer.as_ptr() as _, self.len) }
// ...
\end{lstlisting}

\subsection{Comparison with Existing Tools} \label{evaluation:comparison}
To the best of our knowledge, \TN is the first bug detection tool to systematically detect \bugs in Rust, 
so we are unable to find similar tools to perform an {\it apple-to-apple} comparison. We choose the tools that are able to partially detect \bugs for comparison: Clippy\cite{clippy} and Rudra\cite{Yechan2021Rudra}. We run them on the packages listed in \autoref{top1k-zerodays} and compare the results. The comparison results are shown in \autoref{comparison}. Additionally, we compare the performance of \TN to the Rust's type system on the existing type confusion bugs in RUSTSEC dataset (see \autoref{rustsec-details}).

\vspace{0.05in}
\noindent{\bf Comparison with Clippy.}
Clippy is a static analysis tool that implements more than 650 types of lints~\cite{Lintsoft25online} to detect common errors in the Rust program. 
Clippy supports 2 types of lints to find the unsound usages of \texttt{as} and \texttt{transmute}. 
First, it can check whether \texttt{as} can lead to a misaligned pointer (\texttt{cast\_ptr\_alignment}\cite{castptralignment}). 
Second, it can check whether \texttt{transmute} occurs between types of different Application Binary Interfaces (ABIs) (\texttt{unsound\_collection\_transmute}~\cite{unsoundcollectiontransmute}). 
The version of Clippy with which we compare is 0.1.72, and \autoref{comparison} reveals that Clippy identifies only {10} relevant bugs, all of which are misalignment issues (Type I) restricted to $Con \rightarrow Con$ ({21} bugs in total).
{Regarding the remaining 11 misalignment bugs, one warning is intentionally suppressed by developers, four arise from overlooking variations in ABIs, and six involve \texttt{transmute}.}
We summarize two main reasons for Clippy's limitation as follows.
Firstly, Clippy fails to identify potential bugs involving generic types, as the two lint checks are only carried out when both the \textit{source type} and \textit{dest type} are concrete. Secondly, Clippy does not have a comprehensive approach to identifying type conversion errors. It is observed that Clippy's checks vary between \texttt{as} and \texttt{transmute}; misalignment checks are applied to \texttt{as} but omitted for \texttt{transmute}. For \btwo bugs, checks are conducted exclusively on \texttt{transmute}.
Furthermore, upon reviewing developer comments, it is observed that developers often disregard the warnings due to their belief that Clippy generates a significant number of false positive cases.

\begin{table}[t]
\centering
\caption{{Comparison with Clippy and Rudra}.}
\footnotesize
\begin{tabular}{c c c c c} 
 \toprule
 Detector & Type I & Type II & Type III & Overall \\ 
 \midrule
 Clippy & {10} & 0 & 0 & {10} \\ 
 Rudra & 0 & 0 & 0 & 0 \\
 \TN & {32} & {24} & {15} & {71} \\
 \bottomrule
\end{tabular}
\label{comparison}
\end{table}

\vspace{0.05in}
\noindent{\bf Comparison with Rudra.} 
Rudra\cite{Yechan2021Rudra} is the bug detector that can be used to capture memory safety bugs from Rust packages. 
We compared \TN with Rudra for two reasons. 
First, Rudra is the state-of-the-art memory safety bug detector, reporting 51.6\% of all memory safety bugs. 
Second, Rudra claims it can find the bugs of uninitialized memory exposure, which is the result of \btwo bugs. 
Our evaluation results show that Rudra can find five bugs of uninitialized memory.
However, none of them occurs in the type conversion process,
which means it is not effective in detecting type confusion bugs.
The main reason is that,
for the patterns of type conversion, 
Rudra only detects the terminators at the MIR level.
Unfortunately, \texttt{transmute} has been translated to statements rather than terminators since 2021. 
Moreover, Rudra implements the dataflow checker of \texttt{transmute} but excludes \texttt{as}. 
As a result, Rudra is not able to detect type confusion bugs effectively.

\vspace{0.05in}
\noindent{\bf Comparison with Rust Type System.}
{
We also conduct the experiment to elaborate if  
the existing Rust type system can effectively detect type confusion bugs in \texttt{unsafe} code regions.
%
We compare the positive cases of type confusion bugs detected by \TN and the Rust type system.
Since Rust considers \texttt{unsafe} keyword to be required for raw pointer dereferences and unsafe APIs, removing \texttt{unsafe} will introduce syntax errors.
For the purpose of calculating the precision, we count these unremovable \texttt{unsafe} as positive bug detections by Rust type system.
If there are multiple \texttt{unsafe} blocks in a single function, we count it as one. In other words, the functions with \texttt{unsafe} but no type confusion bugs being reported should be considered as false positive cases.
%
%
We use all vulnerable files including the 32 existing bugs (\autoref{rustsec-details}) 
as the benchmark and find that both \TN and the type system can detect all the existing type confusion bugs; however, the type system produces 116 false positive cases while \TN only produces 3.}

\noindent{\bf Summary.}
Our evaluation confirms that \TN is the most effective tool to detect type confusion bugs in Rust. 
Existing state-of-the-art Rust bug detection tools such as Clippy and Rudra are not as effective,
since they are not designed for detecting type confusion bugs.  
Our experiment also explains that the current Rust type system is not sufficient to check the unsafe type conversion, highlighting the necessity of \TN.

\subsection{Impacts of Type Confusion Bugs} \label{evaluation:sec-impacts}

{
Memory errors might occur due to type confusion bugs, especially if the target type allows access to memory that the source type cannot reach.
The type confusion bugs discovered by \TN have different security implications. 
Among them, {28} trigger panics, {24} cause uninitialized memory access, {8} lead to out-of-bounds access, {7} construct illegal types, and 4 can generate data race issues}.
Using case studies of the \texttt{pprof} package~\cite{pprof}, a popular Rust-based CPU profiler with 1.3k stars on Github and 159 crates.io dependents, we show the impacts of bugs and how it helps diagnose Rust performance bottlenecks. The bug was reported and confirmed by the developers on Github. 
Besides, we also provide other case studies and corresponding PoCs in \autoref{sec:appendixB}.

\begin{lstlisting}[
language=rust, 
style=lst,
caption=Misaligned bug found in the \texttt{pprof} package.,
label=listing:pprof_bug1,
mathescape=false]
impl<'a, T> Iterator for TempFdArrayIterator<'a, T> {
  type Item = &'a T;
  fn next(&mut self) -> Option<Self::Item> {
    // ...
    let length = self.file_vec.len()/size_of::<T>();
    let ts = unsafe {
      slice::from_raw_parts(
        self.file_vec.as_ptr() as *const T,
        length)
      };

pub fn build(&self) -> Result<Report> {
    let mut hash_map = HashMap::new();
    match self.profiler.write().as_mut() {
      Err(err) => {...}
      Ok(profiler) => {
        profiler.data.try_iter()?
          .for_each(|entry| { .. }
\end{lstlisting}

\vspace{0.05in}
\noindent{\bf \Bone Bugs}.
As shown in Listing \ref{listing:pprof_bug1}, \TN detects a \bone bug (line 8) in the \texttt{next} function implemented on \texttt{TempFdArrayIterator}. 
When the unsafe \texttt{slice::from\_raw\_parts} is called (line 11), it assumes the caller meets safety contracts. The raw pointer must be aligned, non-null, and point to \texttt{length} bytes of initialized values\cite{fromrawparts}. Violating these safety contracts causes a panic. 
One way to invoke the \texttt{next} function is to build reports from a running profiler; it will iterate each entry to process the data and write them into reports (line 12). 
The generic type \texttt{T} in \texttt{TempFdArrayIterator} is decided by the item stored in the \texttt{profiler.data} (line 17), which is \texttt{UnresolvedFrames}. \texttt{UnresolvedFrames} is a representation of an event backtrace, and it is a self-defined \texttt{struct} type with fields of \texttt{u64}, \texttt{usize}, an array of \texttt{u8}. 
Since the \texttt{file\_vec} is aligned to 1 byte, any type that has a larger alignment than 1 byte can lead to the \bone bug and crash here.

Famous Rust-based applications like GreptimeDB\cite{greptimedb} (4.1k stars on GitHub) are affected by this bug. GreptimeDB, a time series database storing logs, events, and CPU usage, crashes when calling pprof (\texttt{report::ReportBuilder::build}) to build reports. This panic can obstruct queries or data writes, causing real-time network monitoring applications to miss identifying high CPU usage, leading to network performance decline.
{For example, given a GreptimeDB-based public network server that provides the interface of event backtrace, the attacker could initialize the data \texttt{T} with the type aligned to 2 bytes, causing the network server panic and rendering it unavailable to users.}
After reviewing related GitHub issues, we traced the 	\texttt{ReportBuilder::build} code pattern and searched on GitHub. 
Over 230 code files show similar patterns and may be affected.

\section{Discussion}

\noindent{\bf Rust vs. C++ on Type Confusion Bugs.}
{
The distinction of compiler features, type systems, and memory management in Rust and C++ lead to different types of \bugs and new challenges in detecting them. First, }while pointer-type conversion is the primary cause of bugs in both languages, the \btwo bug and the \bthree bug mentioned in this paper do not occur in C++. This is because the C++ compiler does not rearrange the memory layout of composite types by default (thus no \textit{unstable\_ty}), and it also does not impose strict requirements on the bit-pattern as Rust does.
From a different angle, \bugs in C++, which consistently occur when downcasting an object from a parent class to a child class, does not exist in Rust.
This is because Rust lacks the object characteristics needed for such issues.
Rust introduces the concept of a trait object~\cite{traitobjects}, comparable to C++ objects, to define shared behaviors. However, since trait objects do not involve inheritance, they prevent object-based \bugs.
{Second, before resolving generic types, extracting traits completely is more difficult in Rust since Rust's traits can be implicitly bounded while C++'s concept must be explicit.
Lastly, Rust's implicit memory management brings convenience to developers by avoiding manual management. However, it actually makes pointer alias analysis harder. To verify whether the pointer is still valid, the detector must ensure whether the memory is automatically dropped or the ownership is transferred}.
\looseness=-1

\vspace{0.05in}
\noindent{\bf Complementing Rust type system.}
{
\TN is necessary since expanding the type system protection from Safe Rust to include \texttt{unsafe} is insufficient, as shown in \autoref{evaluation:comparison}.
The type system's safety assurances are not derived from executing type checks. Rather, Safe Rust prevents the presence of an \textit{invalid type} right from the beginning.
Consider the example of a misaligned pointer: Safe Rust prevents developers from converting a pointer to one aligned with larger byte sizes (refer to Listing~\ref{listing:motivation1}). Due to this stringent rule, the compiler confidently assumes that misaligned pointers are not present in Safe Rust. Consequently, optimizations and code generation within the compiler's backend are carried out based on this assumption.
A pointer generated within \texttt{unsafe} is the only entity capable of circumventing these rules. 
Even if the pointer from \texttt{unsafe} is converted to a reference, the compiler will still treat this reference as reliable. This flawed assumption can result in undefined behavior. 
Therefore, a tool like \TN that aids in identifying an invalid type becomes essential for Rust developers.
}

\vspace{0.05in}
\noindent{\bf Mitigations of Type Confusion Bugs.}
We summarize common ways to fix the bugs detected by \TN.

\noindent\emph{Type I: misalignment bugs.}
Misaligned references are prohibited in safe code, but the safety of misaligned raw pointers depends on access methods. We propose two strategies to prevent misalignment issues before dereferencing: (1) use \texttt{read\_unaligned}~\cite{readunalign} or \texttt{write\_unaligned}~\cite{writeunalign}, which handle misaligned pointers, or (2) create a new aligned pointer. Most functions require aligned pointers; \texttt{read\_unaligned} creates an aligned duplicate by copying data (\texttt{copy\_nonoverlapping}) and casting to \texttt{u8}. Alternatively, developers can manually create a new pointer by adding an offset (\texttt{ptr.add}) to align the address.

\noindent\emph{Type II: inconsistent layout bugs.}
To avoid the \btwo bug, we need to ensure the type's memory layout is consistent and stable. If \textit{dst\_ty} is a primitive type with initialized memory in consecutive bytes, \textit{src\_ty} must not have uninitialized bytes. Although most code avoids inconsistent type conversion, it can occur during generic type conversion. 
%
To prevent bugs in generic types, we can apply trait bounds --- list the types that can be legally converted and implement the trait on them. This ensures callers use only the defined types as parameters. A well-known trait implementing this concept is \texttt{bytemuck::Pod}~\cite{Podinbyt5online}.
For \texttt{struct}-to-\texttt{struct} conversions, developers often wrongly assume stable memory layout. To ensure stability, developers need to annotate types for conversion with \texttt{repr(C)} or \texttt{repr(transparent)}~\cite{Typelayo51online}.

\noindent\emph{Type III: mismatched scope bugs.} 
Such bugs often occur in exposed APIs with generic type conversion. 
Developers should limit types and use trait bounds to restrict conversion and validate values before converting.
Libraries provide \texttt{unsafe} APIs like \texttt{from\_*\_unchecked} (e.g., \texttt{str::from\_utf8\_unchecked}~\cite{strutf8unchecked}) for type conversion, whose safety must be ensured by callers. 
Callers must validate that source type values are appropriate for destination types. 
For instance, \texttt{str::from\_utf8\_unchecked} requires UTF-8 valid input, 
unlike the safe \texttt{std::from\_utf8}~\cite{strutf8}, which checks this. 
Developers often use \texttt{from\_utf8\_unchecked} over \texttt{from\_utf8} to avoid overhead, but safe functions should be used in critical security scenarios.

\section{Limitations and Future Work}
{
As detailed in \autoref{evaluation:fp}, \TN has limitations in accurately interpreting different implementations of developer-enforced checks, leading to false positive cases.
%
Some check patterns are inherently implicit and difficult to formalize. For example, calling size check functions but actually examining padding bytes.
Understanding this intricacy requires a deep contextual insights. 
Additionally, assessing the value within the check condition is particularly challenging when relying solely on static tools. 
Without incorporating dynamic analysis, assuring the accuracy of security checks becomes difficult, especially in large software projects. 
In future work, we aim to integrate symbolic execution in access checks, such as implementing constraints to confirm that a pointer's memory address cannot be a multiple of the type size before \TN flags misalignment issues.
}
{Furthermore, with a path limitation of 1 for interprocedural analysis, TypePulse struggles to grasp the context in extended call chains,
which we aim to address in future work.}

\section{Related Work}
\noindent{\bf Research on \bugs in C++/Javascript.} In other programming languages, there are a large number of existing works focusing on \bugs. 
For instance, C++ supports implicit type conversion which can lead to significant issues; thus, numerous scholars have developed various methods to identify the \bugs \cite{jeon2017hextype, haller2016typesan, duck2018effectivesan, lee2015type}. 
Given that C++ includes diverse type casting abilities and runtime polymorphism,
detectors for such bugs must integrate runtime analysis techniques while also managing performance overhead. 
To enhance performance, \texttt{TypeSan}\cite{haller2016typesan} developed a framework capable of efficiently monitoring memory allocation details. 
There are also several research works focusing on \bugs in Javascript\cite{pradel2015typedevil, 9842686, 7958598}. 
Type-related issues in C++ are deemed more severe as they directly contribute to memory safety problems, whereas Javascript typically operates within constrained settings like web browsers. 
Although Rust is engineered with improved type-safety compared to C++ and Javascript, our research shows that type-related errors can still occur in Rust.

\vspace{0.05in}
\noindent{\bf Research on unsafe Rust.}
A substantial body of research explores how the use of \texttt{unsafe} can compromise the integrity of Rust programs~\cite{Qin2020ReplicationPF, Evans2020IsRU, Xu2020MemorySafetyCC, zhang2022towards, papaevripides2021exploiting, mergendahl2022cross, rivera2021keeping, kirth2022pkru}. 
Xu~et al. analyzed hundreds of memory-safety issues, determining that safety assurances can be violated by \texttt{unsafe} code~\cite{Xu2020MemorySafetyCC}, while other works study how to protect the Rust program~\cite{rivera2021keeping, kirth2022pkru}. 
Additionally, some scholars have investigated both memory-safety and concurrency issues, assessing the effects of eliminating unsafe code~\cite{Qin2020ReplicationPF}. 
\texttt{Rudra}~\cite{Yechan2021Rudra} and \texttt{MirChecker}~\cite{Zhuohua2021MirChecker} are specifically designed to target functions that incorporate unsafe code
and detect memory-safety issues.
Observations from our research also suggest a significant correlation between  \bugs in Rust and the use of unsafe code. \looseness=-1

\section{Conclusion}

In this paper, we develop \TN, the {\it first} static analysis tool for detecting \bugs in Rust. 
\TN focuses on detecting the three most common categories of \bugs --- \bone, \btwo, and \bthree.
\TN detected {71} previously unknown bugs from the top 3,000 Rust packages. 
This number surpasses the number of \bugs documented in the last five years in RustSec,
which shows the effectiveness of \TN.
The identified bugs were reported to the developers, who have confirmed {32} of these issues.
We also compare \TN with existing Rust bug detection tools, 
and perform case studies to demonstrate the security implications of the identified bugs.
\TN will be open-sourced to facilitate future research.

\section*{Acknowledgment}
We thank our shepherd and the reviewers for their insightful
feedback. This work is partially supported by ONR grant N00014-23-1-2122 and the IDIA P3 Faculty Fellowship from George Mason University.

\section*{Open Science}

To promote transparency and reproducibility in our research, the data artifacts of this paper will be made publicly available, including source code, detected bugs, and related github issues we reported. We disclose only those issues that have been acknowledged and resolved by developers. Issues that remain unresolved at the time of writing are not included in detail.
All data is available on Zenodo: \url{https://zenodo.org/records/14750104}.

\section*{Ethics Considerations}
We take ethics seriously in this project.
All Rust repositories we tested in the paper are publicly accessible on Github.
During evaluations of our work, \TN identified several previously unknown type confusion vulnerabilities in widely used software. In each case, we followed a responsible disclosure policy, and reported our discovered vulnerabilities to the developers. We also submitted our findings to the CVE program and the RustSec Advisory Database. 
We did not disclose those issues to anyone else.
All the examples mentioned in the paper are the issues that have been acknowledged and fixed. 
The RustSec IDs issued are:
RUSTSEC-2023-0046, RUSTSEC-2023-0047, RUSTSEC-2024-0408, RUSTSEC-2024-0424, RUSTSEC-2024-0426, RUSTSEC-2024-0431;
The CVE ID is currently in the \texttt{reserved} status at the time of writing and will be released later.



\bibliographystyle{plain}
\bibliography{Reference}
\begin{appendix}

\section{Bugs Due to Generic Type Conversion} \label{sec:appendixB}

We demonstrate these three types of type conversion bugs using an example.
The bugs are discovered from \texttt{lmdb-rs} package~\cite{lmdb-rs2023}, which is a package providing API bindings to the LMDB (Lightning Memory-Mapped Database) library. 
In Listing~\ref{listing:motivation_example}, we showcase an implementation of \texttt{from\_mdb\_value} function defined in the \texttt{FromMdbvalue} trait. 
The primary functionality of this code snippet is to convert a reference of \texttt{MdbValue} into another type \texttt{\$t}.
The type conversion is performed using \texttt{transmute} at line 4, which is included in \texttt{unsafe}. 
The function \texttt{new\_from\_size} (line 9) is used to create a new object \texttt{MdbValue} from the reference of generic type \texttt{T}. 
Therefore, the users of the package can create input for \texttt{from\_mdb\_value} with \texttt{new\_from\_size} function.
All type conversion bugs occur in \texttt{from\_mdb\_value} function because of the problematic type conversion.

\begin{lstlisting}[
language=rust, 
style=lst,
caption=The vulnerabilities in \texttt{lmdb-rs} package~\cite{lmdb-rs2023}.,
label=listing:motivation_example,
mathescape=false]
// lmdb-rs/src/traits.rs
impl FromMdbValue for $t {
  fn from_mdb_value(value: &MdbValue) -> $t {
    unsafe { *transmute(value.get_ref()) }
  }
}
// lmdb-rs/src/core.rs
#[inline]
pub fn new_from_sized<T>(data: &'a T) -> MdbValue<'a> {
  unsafe { MdbValue::new(transmute(data), 
           size_of::<T>())}
}
\end{lstlisting}

\begin{lstlisting}[
language=rust, 
style=lst,
caption=Exploit that trigger bug type I.,
label=listing:exploit_bug1,
mathescape=false]
fn main() {
  let a: i32 = 3;
  let mdbval = MdbValue::new_from_sized(&a);
  let res = i64::from_mdb_value(&mdbval);
  println!("{:?}", res);
}
\end{lstlisting}

\vspace{0.05in}
\noindent\textbf{Type I: \Bone bug.} 
The first type of bug occurs when reinterpreting the type of the source object in memory to another type with a larger alignment.
In the exploit code of Listing~\ref{listing:exploit_bug1}, we define and initialize an \texttt{i32} variable \texttt{a}, and convert it into \texttt{i64} using \texttt{from\_mdb\_value} defined in Listing~\ref{listing:motivation_example}, which would cause the misalignment issue.
The data is aligned only if it is stored at an address that is a multiple of the type's alignment bytes. Most primitive types (e.g., u8 and u32 in this case) are aligned to their size. For example, a 32-bit integer (\texttt{i32}) should be stored at the memory address that is a multiple of 4. However, the starting address of \texttt{i32} may not be a multiple of 4, hence accessing a misaligned object can results in undefined behavior. In this case, a misaligned pointer dereference (Line 5 in Listing~\ref{listing:motivation_example}) would cause runtime panic.
Generally, such undefined behaviors in architectures that do not support unaligned access (e.g., before ARMv5) would cause the program to crash.

\vspace{0.05in}
\noindent\textbf{Type II: Inconsistent layout bug.}
In Listing~\ref{listing:exploit_bug2}, the second type of bug occurs when reinterpreting the uninitialized area of the source object in memory. We define a struct \texttt{Padding} (line 3) and instantiate an object \texttt{la} (line 5), and then convert it into an \texttt{i32} primitive object \texttt{res}. It looks like the source and target objects share the same size (\texttt{u8}+\texttt{u16}+\texttt{u8}=\texttt{i32}), but there will be padding bits among each member variable in the struct for alignment. In this case, member \texttt{b}'s size is 16 bits (\texttt{u16}); thus, there will be 8-bit padding for both \texttt{a} and \texttt{c}. These paddings are uninitialized areas in memory, which would trigger the undefined behavior when \texttt{transmute()} accesses them. 
Besides, a further dangerous issue is the unknown padding layout in Rust. Different from the struct padding rule in C (i.e., \texttt{repr(c)}), which usually adds padding bits at the end of the struct, Rust has no guarantees of data layout made by the default representation (\texttt{repr(rust)}). That means the compiler can do whatever it wants to reorder fields based on access patterns. A possible rule in practice is to organize by field size to minimize padding. Therefore, the location of padding is random and may cause data exposure.

\begin{lstlisting}[
language=rust, 
style=lst,
caption=Exploit that trigger bug type II.,
label=listing:exploit_bug2,
mathescape=false]
#[repr(align(2))]
#[derive(Copy, Clone, Debug)]
struct Padding { a: u8, b: u16, c: u8 }
fn main() {
  let la = Padding { a: 10, b: 11, c: 12 };
  let mdbval = MdbValue::new_from_sized(&la);
  let res = i32::from_mdb_value(&mdbval);
  println!("{:?}", res);
}
\end{lstlisting}

\begin{lstlisting}[
language=rust, 
style=lst,
caption=Exploit that trigger bug type III.,
label=listing:exploit_bug3,
mathescape=false]
fn main() {
  let a: i32 = 3;
  let mdbval = MdbValue::new_from_sized(&a);
  let res = bool::from_mdb_value(&mdbval);
  println!("{:?}", res);  // illegal boolean type

  let arr = [1u8; 2];
  println!("{:?}", arr[res as usize]); // OOB index
}
\end{lstlisting}

\vspace{0.05in}
\noindent\textbf{Type III: Mismatched scope bug.} The third bug type happens when the value of the source object exceeds the bit-pattern range of the target type. The bit pattern refers to the raw binary representation of data in memory. In the case of Listing~\ref{listing:exploit_bug3}, we convert an \texttt{i32} variable into the \texttt{bool} type. However, the \texttt{i32} has $2^{32}$ bit-patterns while the boolean type has only 2 bit-patterns (\texttt{false/true}). The value \texttt{false} has the bit pattern \texttt{0x00} and the value \texttt{true} has the bit pattern \texttt{0x01}. Hence, an undefined behavior would occur if an \texttt{bool} object represents any other bit pattern.
Moreover, even the conversion between same-sized types may suffer such an issue. For example, the \texttt{string} type in Rust only supports UTF-8 encoding that includes $(2^8-2)$ unit characters. When we convert an \texttt{u8} ($2^8$ bit-patterns) into \texttt{string} type, the undefined behavior can also be triggered if the value of source object is 254 or 255. 
The third bug can also be exploited to trigger the Out-Of-Bound memory access (OOB). Originally, the compiler always inserts the bound check to protect us from the OOB vulnerability. When the compiler assumes that type has legal value and removes the unnecessary bound check, OOB can be triggered (see line 8).

\section{Root Cause of Type Confusion Bugs} \label{evaluation:rootcause}
Based on \autoref{top1k-zerodays}, we summarize the root cause of \bugs we have found in two phases: First, we discuss how developers make mistakes based on type conversion patterns. 
Second, we study the error-prone methods of {\bf \BC{1}} Conversion and {\bf \BC{2}} Access, specifically on usages of \texttt{unsafe} functions.

\vspace{0.05in}
\noindent{\bf Con $\rightarrow$ Con}.
{{TypePulse identifies more type confusion bugs in the concrete type conversion from the top 3,000 packages.} In concrete type conversion, we highlight the causes of \bone bugs since its number ({21}) is much more than the others ({10} on \btwo bugs and {6} on \bthree bugs). 
We consider the root cause to be the lack of alignment awareness. 
We can also find that developers 
suppress the warnings of alignment from Clippy\cite{clippy}.
While some developers consider the impacts of misalignment are minor since most operating systems nowadays can tolerate the unaligned memory access, we have discovered an issue that can cause to crash (see \autoref{evaluation:sec-impacts})}.

\vspace{0.05in}
\noindent{\bf Gen $\rightarrow$ Con.}
{ In the type conversion $Gen \rightarrow Con$, we have discovered more \btwo bugs (15) than the others (2 on \bone bugs and {4} on \bthree bugs). 
Based on our observation, the developers usually consider the input types that initialize the generic type have a stable memory layout and consequently initialized.
For example, the function \texttt{as\_byte\_slice} is always used to convert the generic type into the slice of \texttt{u8}, leading to uninitialized memory exposure}.

\vspace{0.05in}
\noindent{\bf Con $\rightarrow$ Gen.}
{For bugs related to type conversion $Con \rightarrow Gen$, we find that developers have tried to limit the input types by adding the size check, ensuring the memory layout to be stable. However, the size check is not sufficient to check the alignment and the validity of types.
Nevertheless, we still consider that the developers of the top 3,000 packages provide more protection in this type conversion, leading to the least number ({14}) of bugs compared to the others ({37} in Con $\rightarrow$ Con and {21} in Gen $\rightarrow$ Con)}. 

\vspace{0.05in}
\noindent{\bf \BC{1} Conversion.}
For the methods used for type conversion, we find that developers make more mistakes with \texttt{as} than \texttt{transmute}. 
We assume that developers tend to use \texttt{as} more commonly since \texttt{transmute} is a \texttt{unsafe} function itself while \texttt{as} is not, 
but they are not aware that \texttt{as} can also create 
problematic types,  
even if it is a \texttt{safe} function.
Since we find fewer numbers of \bthree bugs than the other two types of bugs, we consider that the maintainers of the top 3,000 packages are more experienced in avoiding this kind of bug. To support our conjecture, we randomly pick 10 more packages that are not ranked in the top 3,000 and find 6 more \bthree bugs. The bug discovered in \texttt{lmdb-rs} package is one of the examples (see~\autoref{sec:appendixB}).

\vspace{0.05in}
\noindent{\bf \BC{2} Access.}
We also study the \texttt{unsafe} usages of problematic types which could trigger the bugs, and separate them into three categories. 
First, type conversion can cause bugs when developers try to build a slice or vector with unsafe functions such as \texttt{from\_raw\_parts}. 
Second, \textit{dest} is a raw pointer type, and the developers try to dereference the raw pointer. The purpose of raw pointer dereference can be separated into two kinds: a) overwrite the value stored at the memory address and b) dereference the raw pointer to rebuild the reference. 
Third, developers try to use \texttt{transmute} between references, which is dangerous and might break the safety guarantee of reference.

\end{appendix}

\end{document}